\documentclass[aos,preprint]{imsart}

\RequirePackage[OT1]{fontenc}
\RequirePackage{amsthm,amsmath}
\RequirePackage[numbers]{natbib}
\RequirePackage[colorlinks,citecolor=blue,urlcolor=blue]{hyperref}

\arxiv{arXiv:0000.0000}
 
\startlocaldefs
\numberwithin{equation}{section}
\theoremstyle{plain}

\endlocaldefs

\usepackage{amsmath,amssymb,graphicx,xcolor,amsthm}
\usepackage{bbm}
\usepackage{bm}
\usepackage{url}
\usepackage{cancel}
\usepackage[normalem]{ulem}
\usepackage{natbib}
\usepackage{algpseudocode,float,algorithm}
\usepackage{multirow}
\usepackage[normalem]{ulem}
\numberwithin{equation}{section}
\newtheorem{theorem}{Theorem}[section]
\newtheorem{lemma}{Lemma}[section]

{
\theoremstyle{definition}

\newtheorem{remark}{Remark}

}

\def\bmu{\boldsymbol{\mu}}
\def\bdel{\boldsymbol{\delta}}

\def\bbeta{\boldsymbol{\beta}}
\def\bth{\boldsymbol{\theta}}

\DeclareMathOperator*{\argmin}{arg\,min}

\def\diag{{\rm diag}}

\def\supp{{\rm supp}}

\def\R{{\mathbb R}}
\def\tr{{\rm tr}}

\def\E{\mathbb{E}}

\def\1{\mathbbm{1}}
\def\vect{\text{vec}}
\def\Pro{\mathbb{P}}

\def\be{\begin{equation}}
\def\ee{\end{equation}}

\begin{document}

\begin{frontmatter}
\title{A Convex Optimization Approach to High-dimensional Sparse Quadratic Discriminant Analysis\thanksref{T1}}
\runtitle{Sparse QDA}
\thankstext{T1}{The research was supported in part by NSF Grant DMS-1712735 and NIH Grant R01 GM-123056.}

\begin{aug}
\author{\fnms{T. Tony} \snm{Cai}\ead[label=e1]{tcai@wharton.upenn.edu}},
\and
\author{\fnms{Linjun} \snm{Zhang}\ead[label=e2]{linjunz@wharton.upenn.edu}
\ead[label=u1,url]{URL: http://www-stat.wharton.upenn.edu/$\sim$tcai/}}
\runauthor{T. T. Cai and L. Zhang}
\affiliation{University of Pennsylvania}
\address{DEPARTMENT OF STATISTICS\\
THE WHARTON SCHOOL\\
UNIVERSITY OF PENNSYLVANIA\\
PHILADELPHIA, PENNSYLVANIA 19104\\
USA\\
\printead{e1}\\
\phantom{E-mail:\ }\printead*{e2}\\
\printead*{u1}\phantom{URL:\ }\\
}

\end{aug}

%\title[CHIME]{CHIME:  Clustering of High-Dimensional Gaussian Mixtures with {EM} Algorithm and Its Optimality}
%\author[T.T. Cai, J. Ma, \& L. Zhang]{T. Tony Cai, Jing Ma and Linjun Zhang}
% \address{University of Pennsylvania, Philadelphia, USA}

%\author{T. Tony Cai}
%\address{Department of Statistics, University of Pennsylvania}
%\author{Jing Ma}
%\address{Department of Statistics, University of Pennsylvania}
% \author[T. T. Cai, J. Ma and L. Zhang ]{Linjun Zhang}
% \address{Department of Statistics, University of Pennsylvania}

%\author[Author 1 {\it et al.}]{Author 1}
%\author{Author 2}
%\address{Affiliation,
%         City,
%         Country.}
%\maketitle

%%%%%%%%%%%%%%%%%%%%%%%%%%%%%%%%%%%%%%%%%%%%%%%%%%%%%%%%%%%%%%%%
\begin{abstract}
In this paper, we study  high-dimensional sparse Quadratic Discriminant Analysis (QDA) and aim to establish the optimal convergence rates for the classification error. Minimax lower bounds are established to demonstrate the necessity of structural assumptions such as sparsity conditions on the discriminating direction and differential graph for the possible construction of consistent high-dimensional QDA rules. 

We then propose a classification algorithm called SDAR using constrained convex optimization under the sparsity assumptions. Both minimax upper and lower bounds are obtained and this classification rule is shown to be simultaneously rate optimal over a collection of parameter spaces, up to a logarithmic factor.  Simulation studies demonstrate that SDAR performs well numerically. The algorithm is also illustrated through an analysis of prostate cancer data and colon tissue data. The methodology and theory developed for high-dimensional QDA for two groups in the Gaussian setting are also extended to multi-group classification and to classification under the Gaussian copula model.
\end{abstract}



\end{frontmatter}

%%
%%%%%%%%%%%%%%%%%%%%%%%%%%%%%%%%%%%%%%
\section{Introduction}
\label{sec:intro}
%%%%%%%%%%%%%%%%%%%%%%%%%%%%%%%%%%%%%%
 
Discriminant analysis is one of the most commonly used classification techniques in statistics and machine learning due to its simplicity and effectiveness. {Such simplicity mitigates the overfitting when the data has a low dimensional structure, and therefore discriminant analysis has served as a benchmark for} a wide range of applications, including, for example,  face recognition \citep{wright2009robust,rahim2018research, ye2018l1,ju2019probabilistic}, text mining \citep{berry2004survey,abuzeina2018employing}, business forecasting \citep{churchill2006marketing, inam2018forecasting} and gene expression analysis \citep{jombart2010discriminant, li2018libpls, kocchan2019qtqda}.   
In the ideal setting of two known normal distributions $N_p(\bmu_1,\Sigma_1)$ (class 1) and $N_p(\bmu_2, \Sigma_2)$ (class 2), the goal  of the discriminant analysis is to classify a new observation $\bm z$, which is drawn from one of the two distributions with prior probabilities $\pi_1$ and $\pi_2$ respectively, into one of the two classes. In the ideal setting where all the parameters $\bth=(\pi_1,\pi_2,\bmu_1,\bmu_2,\Sigma_1, \Sigma_2)$ are known, the optimal classifier is the quadratic discriminant rule is given by 
\begin{equation}\label{e:intro:qda}
G^*_{\bth}(\bm z)  =  
\begin{cases}
1, \quad  (\bm z-\bmu_1)^\top D(\bm z-\bmu_1)-2\bdel^\top\Omega_2(\bm z-\bar \bmu)-\log({|\Sigma_1|\over |\Sigma_2|})+2\log(\frac{\pi_1}{\pi_2}) >0\\
2, \quad  (\bm z-\bmu_1)^\top D(\bm z-\bmu_1)-2\bdel^\top\Omega_2(\bm z-\bar \bmu)-\log({|\Sigma_1|\over |\Sigma_2|})+2\log(\frac{\pi_1}{\pi_2}) \le0,
\end{cases} 
\end{equation} 
 where $\bm\delta=\bmu_2-\bmu_1$, $\bar \bmu =\frac{\bmu_1+\bmu_2}{2}$, and $D=\Omega_2-\Omega_1$ with $\Omega_i=\Sigma_i^{-1}$ for $i=1, \ 2$, see, for example,  \citet{Anderson2003}.  When $\Sigma_1=\Sigma_2$, the quadratic classification boundary in \eqref{e:intro:qda} becomes linear, reducing the quadratic discriminant analysis (QDA) to the linear discriminant analysis (LDA).  

QDA has been an important technique for classification and is more flexible than the LDA \citep{hastie2009elements}. 
%In particular, this discriminant analysis is a Bayes rule and thus has the minimal classification error. The Bayes classification risk will be used as an oracle benchmark in later discussions. 
 In practice, the parameters $\pi_1,\pi_2,\bmu_1,\bmu_2,\Sigma_1$ and $\Sigma_2$ are usually unknown and instead one observes two independent random samples, $\bm X^{(1)}_1,...,\bm X^{(1)}_{n_1}\stackrel{i.i.d.}{\sim} N_p(\bmu_1,\Sigma_1)$ and $\bm X^{(2)}_1, ..., \bm X^{(2)}_{n_2} \stackrel{i.i.d.}{\sim} N_p(\bmu_2,\Sigma_2)$.  It is practically important to construct a data-driven classification rule based on the two samples. In the low-dimensional setting where the dimension $p$ is small relative to the sample sizes, a natural approach is to simply plug the sample means and sample covariance matrices into the oracle QDA rule \eqref{e:intro:qda}. This approach has been well studied. See, for example, \citet{Anderson2003}. Thanks to the explosive growth of big data, high-dimensional data, where the dimension $p$ can be much larger than the sample sizes, are now routinely collected in scientific investigations in a wide range of fields. %Examples include imaging classification, gene expression arrays, functional magnetic resonance imaging, and web search problems. enlighted by \citet{bickel2004some} and
 In such settings, the conventional LDA and QDA rules perform poorly.

%Considering the high-dimensional LDA problem, to overcome this curse of dimensionality, many high-dimensional generalizations of LDA have been proposed in recent years, and regularity conditions on parameters are needed to ensure that they can be  consistently estimated. One commonly used structural assumption is that both $\Sigma$ (or $\Omega=\Sigma^{-1}$) and $\bdel$ are sparse. Under such assumptions, $\Sigma$ (or $\Omega$) and  $\bdel$ are estimated separately and are then plugged into the oracle LDA, i.e. $G^*_{\bth}(\bm z) $ in \eqref{e:intro:qda} with $\Sigma_1=\Sigma_2$. \citet{shao2011sparse} used the thresholding procedures to estimate $\Sigma$ and $\bdel$, \citet{rothman2008sparse} used the Glasso estimator for $\Omega$.  Another structural assumption was considered in \citet{cai2011direct} and \citet{mai2012direct}. It was observed that the above sparsity assumptions are not made directly on the key quantities needed in the discriminant function, and the classification still performs well even when $\Sigma$ (or $\Omega$) and $\bdel$ cannot be estimated consistently. By assuming the sparsity on the discriminating direction $\bbeta=\Omega\bdel$,  \citet{cai2011direct} and \citet{mai2012direct} estimate $\bbeta$ directly and obtain consistent classification rules. 

%In the case of equal covariance matrices $\Sigma_1=\Sigma_2$, 
For high-dimensional LDA, there already exist a number of proposals and theoretical studies. In particular, assuming sparsity on the discriminating direction, direct estimation methods have been introduced in \citet{cai2011direct} and \citet{mai2012direct} and optimality theory is developed in \citet{Cai2018AdaLDA}.
In contrast, relatively few methods have been introduced for regularized QDA  in the high-dimensional setting and developing an optimality theory is technically more challenging. \citet{li2015sparse} studied high-dimensional QDA by imposing sparsity assumptions on $\bdel$, $\Sigma_1$, $\Sigma_2$ and $\Sigma_1-\Sigma_2$ separately,  and then plugging the estimates of these quantities into the oracle QDA rule \eqref{e:intro:qda}.   \citet{jiang2015quda} introduced a direct estimation approach by assuming that $\Omega_1-\Omega_2$ and $(\Omega_1+\Omega_2)\bdel$ are sparse, and proposed a consistent classification rule. % However, both Theorem 3 in  \citet{li2015sparse} and Theorem 4 in \citet{jiang2015quda}
%\footnote{Although in Corollary 3 of \citet{jiang2015quda}, the authors showed an explicit convergence rate for the classification error of order $s^4\sqrt{\frac{\log p}{n}}$, this result is based on the assumption that the intercept $\eta$ is known.  {\red What's $\eta$?} Theorem 4 in  \citet{jiang2015quda} shows a consistent estimation of $\eta$ without explicit convergence rate. In contrast, our paper shows that even without the assumption of knowing $\eta$, the convergence rate $O(\frac{s\log p}{n}\cdot\log^2 n)$ is achievable. {\red This footnote is long and complicated. It might be better to move it as a remark in a later section.}} 
 %only show the consistency of their proposed classification rules instead of explicit convergence rates. 
 However, it is unclear whether any of these methods achieves the optimal convergence rate for the classification error. %achieve the optimal convergence rates of classification errors.

%Despite the recent progress on discriminant analysis, there has been remarkably little fundamental theoretical study on optimal classification. 
In the present paper, by observing that the oracle rule  (1.1) depends on $\bth$ only through the discriminating direction $\bbeta=\Omega_2\bm \delta$ and differential graph $D=\Omega_2-\Omega_1$, we propose a sparse QDA rule by directly estimating $D$ and $\bbeta$ through convex optimization, and aim to establish the optimality of the proposed classifier  in the high-dimensional settings. It is intuitively clear that QDA is a difficult problem in the high-dimensional setting. For example, it can be seen easily from \eqref{e:intro:qda} that knowledge of the log-determinant of the covariance matrices $\log({|\Sigma_1|\over |\Sigma_2|})$ is essential for the QDA. However, as shown in \citet{Cai2015law}, there is no consistent estimator for the log-determinant of the covariance matrices in the high-dimensional setting even when they are known to be diagonal.
We begin by establishing rigorously minimax lower bound results, which demonstrate that structural assumptions such as sparsity conditions on the discriminating direction $\bbeta$ and differential graph $D$ are necessary for the possible construction of consistent high-dimensional QDA rules.  There are two key steps in obtaining the impossibility results: One is the reduction of the classification error to an alternative loss and another is a careful construction of a collection of least favorable multivariate normal distributions. 

We then propose a classifier called SDAR ({\bf S}parse {\bf D}iscriminant {\bf A}nalysis with {\bf R}egularization) to solve the high-dimensional QDA problem under the sparsity assumptions. The SDAR algorithm proceeds by first estimating $\bbeta$ and $D$ through constrained convex optimization, and then using the estimators to construct a data-driven classification rule. The first estimation step is in a similar spirit to that in \citet{jiang2015quda} by  directly estimating the key quantities in the oracle QDA rule. The second classification step is based on a simple but important observation that $\log(|\Sigma_1|/|\Sigma_2|)=\log(|D\Sigma_1+I_p|)$. As a result, we are able to derive an explicit convergence rate for the classification error of the proposed SDAR algorithm. In addition, we establish a matching minimax lower bound, up to a logarithm factor, that shows the near-optimality of the classifier. Both simulations and real data analysis are carried out to study the numerical performance of the proposed algorithm. The results show that the proposed SDAR algorithm outperforms existing methods in the literature.
The methodology and theory developed for high-dimensional QDA for two groups in the Gaussian setting are also extended to multi-group classification and to classification under the Gaussian copula model.

The contributions of the present paper are three-fold. Firstly, we address the necessity of structural assumptions on the parameters for the  high-dimensional QDA problem by observing that consistent classification is impossible unless $p=o(n)$ without any such assumptions. Secondly, under the sparsity assumptions, we proposed the SDAR rule, and established an explicit convergence rate of classification error. To the best of our knowledge, this is the first explicit convergence rate for high-dimensional QDA. Lastly, we provide a minimax lower bound,  which shows that the convergence rate obtained by the SDAR rule is optimal, up to a logarithmic factor. 

The rest of the paper is organized as follows. In Section \ref{sec:general}, minimax lower bounds are established to show the necessity of imposing structural assumptions for high-dimensional QDA. Section \ref{sec:SDAR} presents in detail the data-driven classification procedure SDAR. Theoretical properties of SDAR are investigated in Section \ref{sec:theory} under certain sparsity conditions. The upper and lower bounds  together show that the SDAR rule achieves the optimal rate for the classification error up to a logarithmic factor.  In Section \ref{non-gaussian}, we consider the semiparametric copula model and introduced a new method called Copula SDAR (CSDAR) and developed corresponding theoretical results for this non-Gaussian model.   Simulation studies are given in Section \ref{sec:simulations} where we compare the performance of the proposed algorithms to other existing classification methods in the literature. In addition, the merits of the SDAR and SDAR classifiers are illustrated through an analysis of a  prostate cancer dataset and a colon tissue dataset. Section \ref{sec:extension} discusses extensions to multi-group classification and to classification under the Gaussian copula model. The proofs of main results are given in Section \ref{sec:proof}, and proofs of other results are provided in the supplement.

%The initialization algorithm, discussion of extensions to the multi-class setting, and the proofs of the main results together with additional technical details, as well as additional simulations are given in the supplement (Cai et al., 2016).}
 
%%%%%%%%%%%%%%%%%%%%
\subsection*{Notation and definitions}
%%%%%%%%%%%%%%%%%%%%

We first introduce basic notation and definitions that will be used throughout the rest of the paper.
For an event $A$, $\1\{A\}$ is the indicator function on $A$. For an integer $m\ge 1$, $[m]$ denotes the set $\{1,2,...,m\}$. Throughout the paper, vectors are denoted by boldface letters. For a vector $\bm u$,  $\|\bm u\|, \|\bm u\|_1, \|\bm u\|_{\infty}$ denotes the $\ell_2$ norm, $\ell_1$ norm, and  $\ell_{\infty}$ norm respectively. We use $\supp(\bm u)$ to denote the support of the vector $\bm u$. $\bm 0_p$ is a $p$-dimensional vector with elements being $0$, and $\bm 1_p$ is a $p$-dimensional vector with elements being $1$. For $i\in[p]$, $\bm e_i$ is the $i$-th standard basis. For a matrix $M\in \R^{p\times p}$, $\|M\|, \|M\|_F, \|M\|_{1}$ denote the spectral norm, Frobenius norm, and matrix $l_1$ norm respectively. In addition, $|M|_1=\sum_{i,j}|M_{i,j}|$, $|M|_\infty=\max_{i,j}|M_{i,j}|$, and  $|M|$ is the determinant of $M$. Let $\lambda_i(M)$ denote the $i$-th eigenvalue of $M$ with $\lambda_1(M)\ge...\ge\lambda_p(M)$.  Let $ M\succ0$ denote M to be a  positive semidefinite matrix and $I_p$ is the $p\times p$ identity matrix. In addition, $M_1\otimes M_2$ denotes the Kronecker product and $\vect(M)$ is the $p^2 \times 1$ vector obtained by stacking the columns of $M$. $\diag(M)$ is the linear operator that sets all the off diagonal elements of M to 0. $E_{i,i}$ is a $p\times p$ matrix whose $(i,i)$-th entry is $1$ and $0$ else. For a positive integer $s < p$, let $\Gamma(s;p)=\{\bm u\in\R^p: \|\bm u_{S^C}\|_1\le \|\bm u_{S}\|_1, \text{ for some }S\subset[p] \text{ with }|S|=s\}$, where $\bm u_{S}$ denotes the subvector of $\bm u$ confined to $S$. For two sequences of positive numbers $a_n$ and $b_n$, $a_n\lesssim b_n$  means that for some constant $c>0$, $a_n \le c\cdot b_n$ for all $n$, and $a_n \asymp b_n$ if $a_n \lesssim b_n$ and $b_n\lesssim a_n$. $a_n\ll b_n$ means that $\lim_{n\to\infty}{|a_n|}/{|b_n|}=0$. In our asymptotic framework, we let $n$ be the driving asymptotic parameter, $s$ and $p$ approach infinity as $n$ grows to infinity. We also use $c, c_1, c_2, ..., C, C_1, C_2$ to denote constants that does not depend on $n,p$, and their values may vary from place to place.  %We write $a_n=\Theta(b_n)$ if $a_n, b_n\to0$ and $a_n\lesssim b_n$ up to logarithm factors. 
%optimal pro- cedures for these two norms are different and consequently matrix estimation un- der the operator norm is fundamentally different from vector estimation. In addi- tion, the results also imply that the banding estimator given in Bickel and Levina (2008a) is sub-optimal under the operator norm and the performance can be sig- nificantly improved.
%%
%%%%%%%%%%%%%%%%%%%%%%%%%%%%%%%%%%%%%%
\section{The Difficulties of High-dimensional QDA}
\label{sec:general}
%%%%%%%%%%%%%%%%%%%%%%%%%%%%%%%%%%%%%%

As mentioned in the introduction, high-dimensional QDA is a difficult problem. In this section, we establish  explicit minimax lower bounds that show the necessity of structural assumptions on the discriminating direction $\bbeta=\Omega_2\bdel$ and differential graph $D=\Omega_2-\Omega_1$ for constructing consistent high-dimensional QDA rules.  

%%%%%%%%%%%%%%%%%%%%
\subsection{The setup}
%%%%%%%%%%%%%%%%%%%%
%{\red Linjun, we need to tie together $n_1$ and $n_2$ with $\pi_1$ and $\pi_2$. In the current form, these have no relationship at all. Then $\pi_1$ and $\pi_2$ cannot be estimated. Same applies to multi-group classification in Section 6.1.}

Suppose we have random samples collected from \\$\pi_1 N_p(\bmu_1, \Sigma_1)+\pi_2 N_p(\bmu_2, \Sigma_2)$, among which $n_1$ samples belong to class 1: $\bm x_1,...,\bm x_{n_1} \stackrel{i.i.d.}{\sim} N_p(\bmu_1,\Sigma_1)$, and $n_2$ samples are in class 2: $\bm y_1, ..., \bm y_{n_2} \stackrel{i.i.d.}{\sim} N_p(\bmu_2, \Sigma_2)$.
The goal is to construct a classification rule $\hat G$, which is a function of $\bm x_i$'s and $\bm y_i$'s, to classify a future data point $\bm z\sim \pi_1 N_p(\bmu_1, \Sigma_1)+\pi_2 N_p(\bmu_2, \Sigma_2)$. This model is parametrized by $\bth=(\pi_1, \pi_2, \bmu_1,\bmu_2,\Sigma_1,\Sigma_2)$.  Let $n=\min\{n_1, n_2\}$. 
For any  classification rule $\hat G:\R^p\to\{1,2\}$, the accuracy is measured by the classification error
\begin{equation}\label{risk}
 R_{\bth}(\hat G)=\E_{\bth}[ \1\{\hat G(\bm z)\neq L(\bm z)  \}],
\end{equation}
where $L(\bm z)$ denotes the true class label of $\bm z$, that is, $L(\bm z)=1$ if $\bm z  \sim N_p(\bmu_1,\Sigma_1)$, and $2$ otherwise.
%{\red (We need to add the following condition in later part: $\frac{n_1}{n_1+n_2}, \pi\in(c_0,1-c_0)$, for some constant $c_0>0$.)}%The performance of the classification rule is measured by the misclassification rate $R(\hat G)=\Pro(\text{label}(\bm x)\neq \hat G(\bm x))$.

When $\bth=(\pi_1, \pi_2, \bmu_1,\bmu_2,\Sigma_1,\Sigma_2)$ is known in advance, the oracle classification rule in \eqref{e:intro:qda} is the Bayes rule and achieves the the minimal classification error, see \citet{Anderson2003}. For ease of presentation, let us define  the discriminant function by
\begin{equation}\label{Qz}
Q(\bm z; \bth)=(\bm z-\bmu_1)^\top D(\bm z-\bmu_1)-2\bdel^\top\Omega_2(\bm z-\bar \bmu)-\log({|\Sigma_1|\over |\Sigma_2|})+2\log(\frac{\pi_1}{\pi_2}).
\end{equation}
Then $Q(\bm z; \bth)=0$ characterizes the classification boundary of the oracle QDA rule, and \eqref{e:intro:qda} can be rewritten as $$
G^*_{\bth}(\bm z) =1+{\1}\{Q(\bm z; \bth)\le0 \},
$$
and $R_{\bth}(G^*_{\bth})=\min_{G\in \mathcal G}R_{\bth}(G)$, where $\mathcal G$ is the set of all classification rules.

%For any classification rule $\hat G$, the classification error is defined by \begin{equation}\label{risk}
% R_{\bth}(\hat G)=\E[ I(\hat G_{\rm SDAR}(\bm z)\neq G(\bm z)  )],
%\end{equation}
In the following the Bayes classification risk $R_{\bth}(G^*_{\bth})$ is used as the benchmark and the excess risk $R_{\bth}(\hat G)-R_{\bth}(G^*_{\bth})$ is used to evaluate the performance of a data-driven classification rule $\hat G$. We say $\hat G$ is consistent, or $G^*_{\bth}$ can be  mimicked by $\hat G$, if the excess risk $R_{\bth}(\hat G)-R_{\bth}(G^*_{\bth})\to 0$
as the sample size $n\to \infty$. 

%%%%%%%%%%%%%%%%%%%%
\subsection{Impossibility of QDA in high dimensions}\label{sec:impqda}
%%%%%%%%%%%%%%%%%%%%

%Recall that our goal is to construct a classification rule based on two groups of $i.i.d.$ observations $\bm x_1,...,\bm x_{n_1} \sim N_p(\bmu_1,\Sigma_1)$, $\bm y_1, ..., \bm y_{n_2}\sim N_p(\bmu_2, \Sigma_2)$. Here we consider the parameter space
%\begin{align}\label{paraspace}
% \Theta_{p}=\{\bth=(\bmu_1, \bmu_2,\Sigma_1, \Sigma_2): \bmu_1, \bmu_2\in\R^p, \Sigma_1, \Sigma_2\in \mathcal D_p\},% \sup_{|x|<\delta} f_{Q,\bth}(x)\le M. \},
%\end{align} 
%where $\mathcal D_p = \{\Sigma\in S^+_p : \Sigma =\diag(\Sigma) \}$ denotes the class of diagonal covariance matrices. %and recall that $D=\Omega_2-\Omega_1$. %$m_1,\delta, M>0$ are three constants. 
We now characterize the fundamental limits of QDA by showing that, without structural assumptions, $G^*_{\bth}$ cannot be  mimicked unless $p\ll n$, which precludes the framework in the high-dimensional settings that motivates our study. %In this section, we begin by observing that when  $\Sigma_1 \neq \Sigma_2$, consistent classification is impossible unless $ p = o(n)$, which precludes the high-dimensional framework that motivates our study.  

We first consider the simple case where $\Sigma_1 = \Sigma_2 = \Sigma$, and in which case the QDA is reduced to the LDA problem. Under the LDA model in the high-dimensional regime, \citet{bickel2004some} and \citet{Cai2018CHIME} proposed consistent classification rules under stringent structural conditions on $(\bmu_1, \bmu_2, \Sigma)$.
%, but they don't have theoretical results showing the necessity of these structural assumptions. 
In this paper, we demonstrate the  the necessity of these structural assumptions by showing that without structural assumptions, a consistent classification rule is impossible in the high-dimensional LDA problem. 

We firstly consider the parameter space $$
\Theta^{(1)}_{p}=\{\bth=({1}/{2},{1}/{2},\bmu_1, \bmu_2, I_p, I_p): \bmu_1, \bmu_2\in\R^p, c_1\le\|\bmu_1-\bmu_2\|\le c_2\},$$ for some constant $c_1, c_2>0$.
%The first impossibility result imply that even when  $\Sigma$ is known to be the identity matrix, $G^*_{\bth}$ cannot be  mimicked without structural assumptions on $\bmu_1, \bmu_2$. %Theorem 4.2 in \citet{Cai2018CHIME}
\begin{theorem}\label{general-lb1}
Suppose that $\hat G$ is any classification rule constructed based on the observations $\bm x_1,...,\bm x_{n} \stackrel{i.i.d.}{\sim} N_p(\bmu_1,I_p)$, $\bm y_1, ..., \bm y_{n} \stackrel{i.i.d.}{\sim} N_p(\bmu_2, I_p)$ with $\bth=({1}/{2},{1}/{2},\bmu_1, \bmu_2, I_p, I_p)\in \Theta^{(1)}_{p}$,  then when $n$ is sufficiently large,  
$$
\inf_{\hat G}\sup_{\bth\in\Theta^{(1)}_p}\E\left[R_{\bth}(\hat G)-R_{\bth}(G^*_{\bth})\right]\gtrsim {\frac{ p}{n}}\wedge 1.
$$ 
\end{theorem}

This theorem implies that even when the covariance matrices are equal and known to be identity matrices, as long as the mean vectors $\bmu_1, \bmu_2$ are unknown, no data-driven method is able to mimic $G^*_{\bth}$ in the high dimensional setting where $ p \gtrsim n$. Structural assumptions are $\bmu_1$ and $\bmu_2$ are necessary for a consistent classification rule. 
%{\red(We may need to add that $\bmu_1$ and $\bmu_2$ can be parted away, that is, $SNR\ge m$. Same for Theorem 2, an upper bound of SNR is also needed.)}

However,  for high-dimensional QDA, structural assumptions on $\bmu_1$ and $\bmu_2$ are not enough and more assumptions are needed. To this end, we consider another scenario where $\bmu_1$ and $\bmu_2$ are known exactly. Let  $\bmu_1^*, \bmu_2^*\in \R^p$  be two given vectors and define the parameter space $$
\Theta^{(2)}_{p}(\bmu_1^*, \bmu_2^*)=\{\bth=({1}/{2},{1}/{2},\bmu_1^*, \bmu_2^*, \Sigma_1, \Sigma_2): \text{$\Sigma_1, \Sigma_2$ are diagonal matrices}\}.
$$ 
%{\red What's the meaning of $\diag(\Sigma_1)=\Sigma_1,\diag(\Sigma_2)=\Sigma_2$? Both are diagonal? If so, it's better to write $\Sigma_1,$ and $\Sigma_2$ are diagonal positive definite matrices.}
 %For the random variable $Q(\bm z; \bth)$ defined in \eqref{Qz}, denote its density function by $f_{Q;\bth}(\cdot)$. 

\begin{theorem}\label{general-lb2}
Suppose ${\hat G}$ is constructed based on the observations $\bm x_1,...,\bm x_{n}$ $\stackrel{i.i.d.}{\sim} N_p(\bmu_1,\Sigma_1)$, $\bm y_1, ..., \bm y_{n} \stackrel{i.i.d.}{\sim} N_p(\bmu_2, \Sigma_2)$. For any given $\bmu_1^*,\bmu_2^*\in \R^p$ with $\|\bmu_1^*-\bmu_2^*\|_2\le C$ where $C>0$ is some constant, when $\bth=({1}/{2},{1}/{2},\bmu_1, \bmu_2, \Sigma_1, \Sigma_2)\in\Theta^{(2)}_{p}(\bmu_1^*, \bmu_2^*)$, we have for  sufficiently large $n$, 
$$
\inf_{\hat G }\sup_{\bth\in\Theta_p^{(2)}(\bmu_1^*,\bmu_2^*)}\E\left[R_{\bth}(\hat G)-R_{\bth}(G^*_{\bth})\right]\gtrsim {\frac{ p}{n}}\wedge 1.
$$ 
\end{theorem}

This theorem implies that even if we have the prior information that $\bmu_1, \bmu_2$ are known and $\Sigma_1, \Sigma_2$ are both diagonal, the quadratic discriminant rule $G^*_{\bth}$  cannot be  mimicked consistently if $ p \gtrsim n$. The construction of consistent classification rules requires stronger assumptions. 

The main strategy of these proofs are discussed in Section \ref{sec:lb}, and the detailed proofs of these lower bound results is provided in Section \ref{sec:pf:imqda}. In addition, the lower bounds are tight, up to a logarithmic factor. %, and we will later provide a procedure that has a matching upper bound up to logarithm factors in Section~\ref{sec:general-ub2}. 
 Specifically, by using the techniques similar to that in Theorem~\ref{Rn}, the plug-in classification rule $\hat G$, which is obtained by plugging in sample means and sample covariance matrices in \eqref{e:intro:qda}, satisfies that $R_{\bth}(\hat G)-R_{\bth}(G^*_{\bth})\lesssim  \frac{p\log^2n}{n} \wedge 1$. This result is further discussed in the supplement.

%%

%%%%%%%%%%%%%%%%%%%%%%%%%%%%%%%%%%%%%%
\section{Sparse Quadratic Discriminant Analysis}
\label{sec:SDAR}
%%%%%%%%%%%%%%%%%%%%%%%%%%%%%%%%%%%%%%

%%%%%%%%%%%%%%%%%%%%
%\subsection{Methodology}
%%%%%%%%%%%%%%%%%%%%

The inconsistency results in Theorems \ref{general-lb1} and \ref{general-lb2} imply the  necessity of imposing structural assumptions on both the mean vectors and covariance matrices. In this section, we consider the QDA problem under the assumptions that the discriminating direction {$\bbeta=\Omega_2\bdel$} and the differential graph $D$ are both sparse. This sparsity assumption, according to \eqref{Qz}, implies that the classification boundary of the oracle rule depends only on a small number of features in $\bm z$. It is also worth noting that the differential graph $D$ corresponds to the change of interactions in two different graphs $\Omega_1$ and $\Omega_2$. The problem of interaction selection is important in its own right and has been studied extensively recently in dynamic network analysis under various environmental and experimental conditions, see \citet{bandyopadhyay2010rewiring,zhao2014direct, xia2015testing, hill2016inferring}.

To see that these two sparsity assumptions are sufficent to obtain a consistent estimator for the optimal classification rule $G^*_{\bth}$, we begin by rewriting $Q(\bm z; \bth)$, defined in \eqref{Qz}. Recall that $\bdel=\bmu_2-\bmu_1,  \bar \bmu =\frac{\bmu_1+\bmu_2}{2}, D=\Omega_2-\Omega_1$ and $\bbeta=\Omega_2\bdel$, then \begin{align}
Q(\bm z; \bth)%&(\bm z-\bmu_1)^\top D(\bm z-\bmu_1)-2\bdel^\top\Omega_2(\bm z-\bmu_1)+\bdel^\top\Omega_2\bdel-\log({|\Sigma_1|\over |\Sigma_2|}) \nonumber\\
=&(\bm z-\bmu_1)^\top D(\bm z-\bmu_1)-2\bbeta^\top(\bm z-\bar \bmu)-\log({|\Sigma_1|\over |\Sigma_2|})+2\log(\frac{\pi_1}{\pi_2}) \nonumber\\
=&(\bm z-\bmu_1)^\top D(\bm z-\bmu_1)-2\bbeta^\top(\bm z-\bar \bmu)-\log(|D\Sigma_1+I_p|)+2\log(\frac{\pi_1}{\pi_2}) . \label{observation}
\end{align}
%where $D=\Omega_2-\Sigma_1^{-2}$,  $\bbeta=\Omega_2\bdel=\Omega_2(\bmu_2-\bmu_1)$, and $\bth=(\mu_1,\mu_2,\Sigma_1, \Sigma_2)$.

A simple but essential observation of \eqref{observation} is that the first three quantities in the above oracle QDA rule $G^*_{\bth}$ depends on either $D$ or $\bbeta$, and the forth term $\log(\pi_1/\pi_2)$ is easy to estimate. % {\red Not correct. Also depends on $\log({|\Sigma_1|\over |\Sigma_2|})$, $\Sigma_1$, and $\log(\frac{\pi_1}{\pi_2}) $. We can perhaps say that ``The key quantities in the oracle QDA rule $G^*_{\bth}$ are  $D$ and $\bbeta$."}
In the present paper, we shall show that  under the sparsity assumptions on these two quantities, $D$ and $\bbeta$ can be estimated directly and efficiently, and the classification rule based on these two estimates enjoys desirable theoretical guarantees.

\begin{remark}
{\rm By symmetry, $Q(\bm z;\bth)$ can also be rewritten in a form that depends on $(\Omega_1+\Omega_2)\bdel$ and $D$. The reason that we consider $(\Omega_2\bdel, D)$ as the key quantity is that this could be easily extended to the case with $K$ multiple groups. In this generalized setting, we consider using the first group as a benchmark, and computing the likelihood ratio of other groups versus the first one. As a result, the key quantity in the multiple classification case is $\{(\Omega_k(\bmu_k-\bmu_1), \Omega_k-\Omega_1)]\}_{k=2}^K$. See more discussion in Section~\ref{sec:extension}.
}\end{remark}

In the following, we proceed to estimate $D$ and $\bbeta$ through constrained convex optimization. Let the first sample covariance matrix be $\hat\Sigma_1=n_1^{-1}\sum_{i=1}^{n_1}(\bm x_i-\hat \bmu_1)(\bm x_i-\hat \bmu_1)^\top$, where $\hat\bmu_1=n_1^{-1}\sum_{i=1}^{n_1}\bm x_i$ and define $\hat\Sigma_2$ and $\hat\bmu_2$ similarly. Since $D$ satisfies the equation $\Sigma_1 D\Sigma_2=\Sigma_1-\Sigma_2$ and $\Sigma_2 D\Sigma_1=\Sigma_1-\Sigma_2$, a sensible estimation procedure is to solve $\hat\Sigma_1 D \hat\Sigma_2/2+\hat\Sigma_2 D \hat\Sigma_1/2-\hat\Sigma_1+\hat\Sigma_2=0$ for $D$.  We estimate $D$ through the following constrained $\ell_1$ minimization approach
\begin{equation}
\hat D=\arg\min_{D\in\R^{p\times p}} \left\{|D|_1: \, |\frac{1}{2}\hat\Sigma_1 D \hat\Sigma_2+\frac{1}{2}\hat\Sigma_2 D \hat\Sigma_1-\hat\Sigma_1+\hat\Sigma_2|_{\infty}\le\lambda_{1,n}\right\}
\label{opt1},
\end{equation}
where $\lambda_{1,n}=c_1\sqrt{\frac{\log p}{n}}$ is a tuning parameter with some constant $c_1>0$ that will be specified later. 

%To solve \eqref{opt1}, let's transform it to an equivalent optimization problem
%\begin{align}
%\hat D=&\arg\min_{D\in\R^{p\times p}} |D|_1\nonumber\\
%&\text{subject to }\|(\frac{1}{2}\hat\Sigma_1 \otimes \hat\Sigma_2+\frac{1}{2}\hat\Sigma_2 \otimes \hat\Sigma_1)\vect(D)-\vect(\hat\Sigma_1)+\vect(\hat\Sigma_2)\|_{\infty}\le\lambda_{1,n}\label{opt1v2}.
%\end{align}
\begin{remark}{\rm
The estimator $\hat D$ defined in \eqref{opt1} is similar to that in \citet{zhao2014direct}, but has better numerical performance due to symmetrization. In addition, we are able to solve \eqref{opt1} in a more computationally efficient way. % than that in \citet{zhao2014direct}. 
\citet{zhao2014direct} vectorized $D$ and transformed the optimization problem \eqref{opt1} to a linear programming with a $p^2 \times p^2$ constraint matrix $\hat\Sigma_1\otimes\hat\Sigma_2$, which is computationally demanding for large $p$. In contrast, we solve \eqref{opt1} by using the primal-dual interior point method \citep{candes2005l1}, and keep the matrix form of $D$ in each step of conjugate gradient descent, by using the matrix multiplications $\frac{1}{2}\hat\Sigma_1 D \hat\Sigma_2+\frac{1}{2}\hat\Sigma_2 D \hat\Sigma_1$ instead of computing $(\frac{1}{2}\hat\Sigma_1 \otimes \hat\Sigma_2+\frac{1}{2}\hat\Sigma_2 \otimes \hat\Sigma_1)\vect(D)$ repeatedly. As a result, the computational complexity is reduced to $O(p^3)$ from $O(p^4)$, and our method is able to handle the problem with larger dimension $p$. The code is available at \url{https://github.com/linjunz/SDAR}.
}\end{remark}
% Therefore, when we have the prior knowledge that both $\Omega_1$ and $\Omega_2$ are sparse, we use CLIME in \citet{cai2011constrained} to estimate $\Omega_1$ and $\Omega_2$ separately, and then use their difference to estimate the differential matrix $D$. In addition, a speed modified algorithm to solve \eqref{opt1v2} can be found in \citet{zhao2014direct}.}

We then proceed to estimating $\bbeta$. Similarly, since the true $\bbeta$ satisfies that $\Sigma_2\bbeta=\bmu_2-\bmu_1$, following \citet{cai2011direct}, $\bbeta$ can be estimated by the following procedure 
\begin{equation}
\hat\bbeta=\arg\min_{\bbeta\in\R^{p}}\left\{\|\bbeta\|_1: \, \| \hat\Sigma_2 \bbeta -\hat\bmu_2+\hat\bmu_1\|_{\infty}\le\lambda_{2,n}\right\}
\label{opt2},
\end{equation}
where $\lambda_{2,n}=c_2\sqrt{\frac{\log p}{n}}$ is a tuning parameter with some constant $c_2>0$.

We estimate $\pi_1$ and $\pi_2$ by $\hat\pi_1=\frac{n_1}{n_1+n_2}$ and $ \hat\pi_2=\frac{n_2}{n_1+n_2}$ respectively. Given the solutions $\hat D$ and $\hat \bbeta$ to \eqref{opt1} and \eqref{opt2} and the estimates $\hat \pi_1$ and $\hat \pi_2$,  we then propose the following classification rule: classify $\bm z$ to class $1$ if and and only if 
$$
(\bm z-\hat\bmu_1)^\top \hat D(\bm z-\hat\bmu_1)-2\hat\bbeta^\top(\bm z-\frac{\hat\bmu_1+\hat\bmu_2}{2})-\log(|\hat D\hat\Sigma_1+I_p|)+\log(\frac{\hat\pi_1}{\hat\pi_2}) >0.
$$
We shall call this rule the Sparse quadratic Discriminant Analysis rule with Regularization (SDAR), and denote it by $\hat G_{\rm SDAR}$. Analytically, it's written as \begin{align}\label{SDAR}
&\hat G_{\rm SDAR}(\bm z)=1+\\
\notag&{\1}\{(\bm z-\hat\bmu_1)^\top \hat D(\bm z-\hat\bmu_1)-2\hat\bbeta^\top(\bm z-\frac{\hat\bmu_1+\hat\bmu_2}{2})-\log(|\hat D\hat\Sigma_1+I_p|)+\log(\frac{\hat\pi_1}{\hat\pi_2}) \le 0\}.
\end{align}

The SDAR rule is easy to implement as  both $\eqref{opt1}$ and $\eqref{opt2}$ can be solved by linear programming. We shall show in the next sections that the SDAR rule has desirable properties both theoretically and numerically.

%%%%%%%%%%%%%%%%%%%%
\section{Theoretical Guarantees}
\label{sec:theory}
%%%%%%%%%%%%%%%%%%%%

We now study the accuracy of the estimators $\hat D$ and $\hat \bbeta$ in \eqref{opt1} and \eqref{opt2}, and the performance of the  resulting classifier $\hat G_{\rm SDAR}$ in \eqref{SDAR}. We first establish the rates of convergence for the estimation and classification error and then provide matching minimax lower bounds, up to logarithm factors. These results together show the near-optimality of the SDAR  rule.

\subsection{Upper bounds}
%%%%%%%%%%%%%%%%%%%%

To overcome the limitations illustrated in Section \ref{sec:general}, we consider the following parameter space of $\bth=(\pi_1,\pi_2, \bmu_1, \bmu_2, \Sigma_1, \Sigma_2)$. Especially,  we assume here that both the discriminating direction $\bbeta$ and the differential graph $D$ are sparse.  Let $f_{Q,\bth}$ be the probability density of $Q(\bm z; \bth)$ defined in \eqref{Qz}, we consider the following parameter space.
\begin{equation}\label{paraspace}
 \begin{aligned}
\Theta_{p}(s_1, s_2)=\{&\bth=(\pi_1,\pi_2,\bmu_1, \bmu_2,\Sigma_1, \Sigma_2): \bmu_1, \bmu_2\in\R^p, \Sigma_1, \Sigma_2\succ 0,|D|_0\le s_1, \|\bbeta\|_0\le s_2 \\
&{\|D\|_F, \|\bbeta\|_2\le M_0}, M_1^{-1}\le\lambda_{\min}(\Sigma_k)\le\lambda_{\max}(\Sigma_k)\le M_1, k=1,2,\\
&{\sup_{|x|<\delta}f_{Q,\bth}(x)< M_2}, c\le\pi_1,\pi_2\le1-c
\},
\end{aligned} 
\end{equation}
for some constants $M_0>0, M_1>1$, $\delta, M_2>0$ and $c\in(0,1/2)$.

%We introduce four conditions as follows:
% \begin{description}
%%\item[(C1)] $\|\Sigma_2\|_2\le C_0$ for some $C_0>1$.
%\item[(C1)] The differential graph $D=\Omega_2-\Omega_1$ is overall sparse, i.e.  $D$ has $s_1$ has non-zeros entries with $s_1$. %\le\sqrt p$. 
%In addition, ${\|D\|_F}\le M$ and, $M_1^{-1}\le\lambda_{\min}(\Sigma_1)\le\lambda_{\max}(\Sigma_1)\le M_1$, and $M_1^{-1}\le\lambda_{\min}(\Sigma_2)\le\lambda_{\max}(\Sigma_2)\le M_1$ for some constant $M_1>1$. 
%\item[(C2)] The discriminant direction $\bbeta=\Omega_2(\bmu_2-\bmu_1)$ is sparse with sparsity $s_2$. ${\|\bbeta\|_2}\le M_3$ for some constants $M_2, M_3>0$.
%{\color{black}\item[(C3)] Let $f_{Q,\bth}$ be the probability density of $Q(\bm z; \bth)$, defined in \eqref{Qz}. $\sup_{|x|<\delta}f_{Q,\bth}(x)< M_4$ for some constant $\delta, M_4>0$.} ({\cyan This condition can be replaced by a relaxed condition: There exists a constant $c_1\in(0,{1})$, such that $\frac{\|D\|}{\|D\|_F}\le c_1$. This can be proved by separating the positive and negative parts. Maybe we can remove this condition completely})
%%\item[(C4)] There exists a constant $c\in(0,\frac{1}{2})$, such that $R_{\bth}(G^*_{\bth})\le c.$
%\end{description}
\begin{remark}{\rm
Note that we assume sparsity on both the discriminant direction $\bbeta$ and the differential graph $D$, whose necessities are shown by Theorem \ref{general-lb1} and \ref{general-lb2}. The upper bound on $\|\bbeta\|_2$ is a general assumption in LDA, see \citet{cai2011direct,neykov2015unified}; and \citet{Cai2018CHIME}, and we assume the same on $\|vec(D)\|_2=\|D\|_F$ in the QDA setting. Moreover, the condition on the bounded density is commonly assumed in discriminant analysis, see the margin assumption in \citet{mammen1999smooth}, condition (C1) in \citet{cai2011direct}, and discussions in \citet{li2015sparse} and \citet{jiang2015quda}. In the following we present a condition on $\bth$ such that this bounded density assumption holds. Note that the term $\bm z^\top D\bm z+\bbeta^\top\bm z$ is equal in distribution to a weighted non-central chi-square distribution, by using the similar proof as that of Lemma 7.2 in \citet{xu20142}, the
condition ${\sup_{|x|<\delta}f_{Q,\bth}(x)< M_2}$ holds when either the two largest positive eigenvalues of $D$ $\lambda_1(D),\lambda_2(D)$ or the two largest negative eigenvalues of $D$ $\tilde\lambda_1(D),\tilde\lambda_2(D)$ are of the same order, that is $0<\liminf_{n\to\infty}\frac{\lambda_1(D)}{\lambda_1(D)+\lambda_2(D)}<\limsup_{n\to\infty}\frac{\lambda_1(D)}{\lambda_1(D)+\lambda_2(D)}<1$ or $0<\liminf_{n\to\infty}\frac{\tilde\lambda_1(D)}{\tilde\lambda_1(D)+\tilde\lambda_2(D)}<$ $\limsup_{n\to\infty}\frac{\tilde\lambda_1(D)}{\tilde\lambda_1(D)+\tilde\lambda_2(D)}<1$.
}\end{remark}
 %{\blue It's not easy to remove the conditions on bounded norms. The problem is on the lower bound result. I will check all the proof again in the next two days. } 

At first, we show that over the parameter space $\Theta_p(s_1,s_2)$, the estimators $\hat D$, $\hat\bbeta$ obtained in \eqref{opt1} and \eqref{opt2} converge to the true parameters $D$ and $\bbeta$. This theorem will then be used to establish the consistency of the proposed classification rule. 

\begin{theorem}\label{estimation}
Consider the parameter space $\Theta_p(s_1, s_2)$, and assume that $n_1\asymp n_2, s_1+s_2\lesssim \frac{{n}}{\log p}$, where $n=\min\{n_1, n_2\}$. In optimization problems  \eqref{opt1} and \eqref{opt2}, let $\lambda_{i,n} = c_i\sqrt{ \log p/n}$ with $c_i > 0$, $i=1,2$ being sufficiently large constants. Then the estimators obtained in    \eqref{opt1} and \eqref{opt2} satisfies that, with probability at least $1-p^{-1}$, 
\begin{align*}
&\|\hat D-D\|_F\lesssim \sqrt{\frac{s_1\log p}{n}}; \quad \|\hat\bbeta-\bbeta\|_2\lesssim \sqrt{\frac{s_2\log p}{n}}.
%&\|\hat D-D\|_F\lesssim \sqrt{\frac{s_1\log p}{n}};\\
%&\|\hat\bbeta-\bbeta\|_2\lesssim \sqrt{\frac{s_2\log p}{n}}.
\end{align*}
\end{theorem}

The above theorem shows that although our estimating procedure \eqref{opt2} is different from \citet{zhao2014direct}, the same convergence rate can be obtained and requires milder theoretical conditions. In fact, \citet{zhao2014direct} assumes that $\|\Omega_1\|_1$ and $\|\Omega_2\|_1$ are both bounded, and additionally requires that the off-diagonal elements of $\Sigma_1$ and $ \Sigma_2$ are vanishing as $n\to\infty$, which is much stronger than conditions in \eqref{paraspace}. %{\cyan[More details later.]} 
%We then proceed to characterize the accuracy of the classification rule $\hat G_{\rm SDAR}$
%To establish the convergence rate, we introduce the following conditions \begin{description}
%%\item[(C1)] $\|\Sigma_2\|_2\le C_0$ for some $C_0>1$.
%\item[(C1)] The differential graph $D=\Omega_2-\Omega_1$ are overall sparse, i.e.  $D$ has $s_1$ has non-zeros entries with $s_1<p$. In addition, $\|\Omega_1\|_1, \|\Omega_2\|_1\le C_1$ for some constant $C_1>0$ and any $k\ge 2$.
%\item[(C2)] The discriminant direction $\bbeta=\Omega_2(\bmu_2-\bmu_1)$ is sparse with sparsity no larger than $s_2<p$. $\|\Sigma_2\|_2\le C_2$, $\|\bmu_k\|_\infty\le C_3$ for some constants $C_2, C_3>0$ and any $k=1,2$.
%\item[(C3)] There exist constants $c_1, c_2>0$, such that $c_1<R_{\bth}(G^*_{\bth})<\frac{1}{2}-c_2.$
%\end{description}
In addition, the above bound implies that when $\Sigma_1=\Sigma_2$, that is, $s_1=0$, we have $\hat D=D=0$ when $\lambda_{1,n}$ is suitably chosen. This implies that when the two covariance matrices are equal, SDAR rule \eqref{SDAR} would adaptively be reduced to the LPD rule in \citet{cai2011direct} designed for high-dimensional LDA.

We now turn to the performance of the classification rule $\hat G_{\rm SDAR}$. The behavior of $\hat G_{\rm SDAR}$ is measured by the excess risk $R_{\bth}(\hat G_{\rm SDAR})-R_{\bth}(G^*_{\bth})$, defined in \eqref{risk}. The following theorem provides the upper bound for the excess classification error.

\begin{theorem}\label{Rn}
Consider the parameter space $\Theta_p(s_1, s_2)$, and assume that $n_1\asymp n_2, s_1+s_2\lesssim \frac{{n}}{\log p\cdot\log^2 n}$. Then the proposed SDAR classification rule in \eqref{SDAR} satisfies that,
$$
\sup_{\bth\in\Theta_p(s_1, s_2)}\E\left[R_{\bth}(\hat G_{\rm SDAR})-R_{\bth}(G^*_{\bth})\right]\lesssim  (s_1+{s_2})\cdot {\frac{\log p}{n}}\cdot\log^2 n.
$$ 
\end{theorem}
The result in Theorem \ref{Rn} shows that $\hat G_{\rm SDAR}$ is able to mimic $G^*_{\bth}$ consistently over the parameter space $\Theta_p(s_1, s_2)$, and to the best of our knowledge, gives the first explicit convergence rate of classification error for the  high-dimensional QDA problem.

\begin{remark}{\rm
Related work studying the convergence of classification error includes \citet{li2015sparse} and  \citet{jiang2015quda}, but both Theorem 3 in  \citet{li2015sparse} and Theorem 4 in \citet{jiang2015quda}  only show the consistency of their proposed classification rules instead of explicit convergence rates.  Although in Corollary 3 of \citet{jiang2015quda}, the authors showed a convergence rate for the classification error of order $s_1s_2^2\sqrt{{\log p}/{n}}$ under some regularity conditions, this result is based on the assumption that an intercept term $\eta$, defined in their paper, %= 2\log(\pi_1 /\pi_2 )+ 1/4(\bmu_1-\bmu_2)^\top\Omega(\bmu_1-\bmu_2) + \log |\Sigma_2 | - \log |\Sigma_1 |$ 
is known.   \citet{jiang2015quda} proposed to estimate $\eta$ based on the idea of cross validation and in their theorem 3 they showed the consistency of this estimation without explicit convergence rate. In contrast, our paper shows that the convergence rate $O({(s_1+s_2)\log p}\cdot\log^2 n/{n})$ is achievable, which is much faster than their results. In addition, the assumptions here are weaker. 
}\end{remark}

% and improves the convergence rate of the classification error of the QUDA rule given in \citet{jiang2015quda}. In fact, \citet{jiang2015quda} shows the convergence rate $\sqrt{s\log p/n}$ under some stronger conditions, which is much slower than the convergence rate $s\log p/n\cdot\log^2n$ shown in Theorem~\ref{Rn}. 
The major technical challenge of this improvement is the characterization of the distribution of $Q({\bm z;\bth})$, which involves the sum of weighted non-central chi-square random variables. In the next section we will show that this convergence rate is indeed optimal up to logarithm factors. 

%{\cyan Comment: In high-dimensional QDA literature,  \citet{li2015sparse} and \citet{jiang2015quda} studied the convergence of $R_{\bth}(G^*_{\bth})$ theoretically.  \citet{li2015sparse} shows $R_{\bth}(\hat G_{\rm SDAR})-R_{\bth}(G^*_{\bth})=o_P(1)$. \citet{jiang2015quda} shows $R_{\bth}(\hat G_{\rm SDAR})-R_{\bth}(G^*_{\bth})=o_P({n^{-1/2}})$, but their proof is wrong.}

%%%%%%%%%%%%%%%%%%%%%%%%%%%%%%%%%%%%%%
\subsection{Minimax lower bound for sparse QDA} 
\label{sec:lb}
%%%%%%%%%%%%%%%%%%%%%%%%%%%%%%%%%%%%%%

In this section we establish the minimax lower bound for the convergence rate of  $R_{\bth}(\hat G)-R_{\bth}(G^*_{\bth})$, and thus show the optimality of $\hat G_{\rm SDAR}$ up to logarithm factors.

%For the random variable $Q(\bm z; \bth)$ defined in \eqref{Qz}, denote its density function by $f_{Q;\bth}(\cdot)$.
% Recall that $D=\Omega_2-\Omega_1$ and $\bbeta=\Omega_2\bdel$, and we are going to establish the minimax lower bound over the following parameter space
% {\cyan\begin{align}
%\notag\Theta_{p}(s_1, s_2)=\{&\bth=(\bmu_1, \bmu_2,\Sigma_1, \Sigma_2): \bmu_1, \bmu_2\in\R^p, \Sigma_1, \Sigma_2\in S^+_p,|D|_0\le s_1, \|\bbeta\|_0\le s_2 \},
%\end{align} 
%}
%The following lower bound shows the optimality of the proposed SDAR rule $\hat G_{\rm SDAR}$.
\begin{theorem}\label{sparse-lb}
Consider the parameter space $\Theta_p(s_1, s_2)$ defined in \eqref{paraspace}. Suppose $n_1\asymp n_2$, $1\le s_1, s_2\le o(\frac{{n}}{\log p})$, and $\hat G$ is constructed based on the observations $\bm x_1,...,\bm x_{n} \stackrel{i.i.d.}{\sim} N_p(\bmu_1,\Sigma_1)$, $\bm y_1, ..., \bm y_{n} \stackrel{i.i.d.}{\sim} N_p(\bmu_2, \Sigma_2)$. Then the minimax risk of the classification error over   $ \Theta_{p}(s_1, s_2)$ satisfies  
$$
\inf_{\hat G}\sup_{\bth\in  \Theta_{p}(s_1, s_2)}\E\left[R_{\bth}(\hat G)-R_{\bth}(G^*_{\bth})\right]\gtrsim (s_1+s_2)\cdot{\frac{\log p}{n}}.
$$ 
\end{theorem}

\begin{remark}
 Theorems \ref{Rn} and \ref{sparse-lb} together show that the proposed SQDA rule is optimal for classifying Gaussian data under mild regularity conditions. No other method can have a faster convergence rate of misclassification error in this region. The method and results can be further extended beyond the Gaussian setting.  See Section~\ref{non-gaussian} for a detailed discussion on the extension.
\end{remark}

%Theorem \ref{general-lb} is a special case of this theorem when we take $s_2=\sqrt p$, and this justifies the necessity to assume the sparsity of the differential graph $D$ when we consider the high-dimensional case where $\sqrt p\ge n$.

The challenge of proving Theorem \ref{sparse-lb} is that the excess risk $R_{\bth}(\hat G)-R_{\bth}(G^*_{\bth})$ does not satisfy the triangle inequality (or subadditivity), which is essential to the standard minimax lower bound techniques. To overcome this challenge, we define an alternative risk function $L_{\bth}(\hat G)$ as follows, 
%Let $G_{QDA; \bth}$ be the optimal rule (Fisher's rule) with parameter $\bth$, and for any classification rule $\hat G$ 
\begin{equation}\label{Ln}
L_{\bth}(\hat G):=\Pro_{\bth}\left(\hat G(\bm z)\neq G^*_{\bth}(\bm z)\right).
\end{equation}

This loss function $L_{\bth}(\hat G)$ is essentially the probability that $\hat G$ produces a different label than $G^*_{\bth}$, and satisfies the triangle inequality, as shown in Lemma~\ref{triangle}. The connection between $R_{\bth}(\hat G)-R_{\bth}(G^*_{\bth})$ and $L_{\bth}(\hat G)$ is presented by the following lemma, which shows that it's sufficient to provide a lower bound for $L_{\bth}(\hat G)$ to prove Theorem \ref{sparse-lb}. 
\begin{lemma}\label{transition} 
 Suppose $\bth \in   \Theta_{p}(s_1, s_2)$. There exists a constant $c>0$, doesn't depend on $n,p$, such that for some classification rule $G$, if $L_{\bth}(G)<c$,
then,
$$L_{\bth}^2(G)\lesssim \Pro_{\bth}(G(\bm z)\neq L(\bm z))-\Pro_{\bth}(G_{\bth}(\bm z) \neq L(\bm z)).$$
\end{lemma}

Based on Lemma \ref{transition}, we use Fano's inequality on a carefully designed least favorable multivariate normal distributions  to complete the proof of Theorems \ref{general-lb2} and \ref{sparse-lb}. The details are shown in Section \ref{sec:proof}.

%%%%%%%%%%%%%%%%%%%%%%%%%%%%%%%%%%%%%%
\section{Extension to the non-Gaussian distributions}\label{non-gaussian}
%%%%%%%%%%%%%%%%%%%%%%%%%%%%%%%%%%%%%%

%%%%%%%%%%%%%%%%%%%%%%%%%%%%%%%%%%%%%%
%\subsection{Classification under Gaussian copula model}
%%%%%%%%%%%%%%%%%%%%%%%%%%%%%%%%%%%%%%

The Gaussianity assumption can be relaxed by incorporating semiparametric Gaussian copula model into the QDA framework. This larger semiparametric Gaussian copula model enables robust estimation and classification, and has been studied widely in statistics and machine learning, including  linear discriminant analysis (LDA) \citep{han2013coda, mai2015sparse}, correlation matrix estimation \citep{han2017statistical}, graphical models \citep{liu2012high, xue2012regularized}, and linear regression \citep{cai2018high}.

 The Semiparametric Discriminant Analysis (SeDA) model, introduced by \citet{lin2003discriminant}, assumes that there are two groups of $p$-dimensional observations $\bm x^{(1)}_1,...,\bm x_{n_1}^{(1)}\sim \bm X^{(1)}$, $\bm x^{(2)}_1,...,\bm x_{n_2}^{(2)}\sim \bm X^{(2)}$, and there are some unknown strictly increasing functions $f_{1}, ..., f_{p}$, such that 
 \begin{equation}\label{model:seda}
 {\bm f}(\bm X^{(k)})\stackrel{def}{=}(f_{1}(X^{(k)}_1), ..., f_{p}(X^{(k)}_p))\sim N_p(\bmu_k, \Sigma_k) \text{ for } k=1,2.
\end{equation}  

By properties of the Gaussian distribution, $f_{j}$'s are only unique up to location and scale shifts. Therefore, for identifiability, same as \citet{mai2015sparse}, we assume, for  $j=1,...,p$ 
\begin{equation}\label{assumption:copula}
\E[f_j(X^{(1)}_j)]=\mu_j^{(1)}=0;\; Var(f_j(X^{(1)}_j))=\sigma_{jj}^{(1)}=1.
\end{equation}

The SeDA model in the high-dimensional LDA setting was recently studied by \citet{mai2015sparse} and \citet{han2013coda} under the assumption that $\Sigma_k$'s are all equal. By applying the LPD idea in \citet{cai2011direct}, consistent classification rules were proposed under this semiparametric linear discriminant analysis model. 

The current paper presents a framework to extend the high-dimensional semiparametric LDA to high-dimensional semiparametric QDA. %Analogous analysis can  be carried out to high-dimensional semiparametric quadratic discriminant analysis to relax the equal covariance matrices assumption. 
Estimating the mean vectors and covariance matrices similarly as in  \citet{mai2015sparse,han2013coda} and then plugging these estimators in \eqref{opt1} and \eqref{opt2} would lead to a generalized classification rule under the semiparametric quadratic discriminant analysis model. Specifically, for $j\in\{0,1,2,...,p \}$, let $\hat F_j^{(k)}(t)$ be the empirical cumulative distribution function of $\{x^{(k)}_{ij}\}_{i=1}^{n_k}$ Winsorized   at $(1/n_k^2,1-1/n_k^2)$ \citep{mai2015sparse}, with  $j\in\{1,2,...,p\}$. In addition,  we estimate the mean $\mu^{(2)}_j$ and variance $\sigma_{jj}^{(2)}$ respectively by \begin{equation}\label{copula-est}
\hat\mu^{(2)}_j=\frac{1}{n_2}\sum_{i=1}^{n_2}\Phi^{-1}\circ\hat F_j^{(1)}(x_{ij}^{(2)}), \hat\sigma_{jj}^{(2)}=\frac{1}{n_2-1}\sum_{i=1}^{n_2}\big(\Phi^{-1}\circ\hat F_1^{(1)}(x_i^{(2)})-\hat\mu^{(2)}_j\big)^2.
\end{equation}

Here we note that by the identifiability assumption \eqref{assumption:copula}, we have $\hat\mu^{(1)}_j=0, \hat\sigma_{jj}^{(1)}=1$ for $j=1,2,...,p$.

Then, we estimate the correlation matrices of $\bm X^{(1)}$ and $\bm X^{(2)}$ the same way as \citet{han2013coda}. For $j_1\neq j_2\in[p]$, we firstly let the Kendall's tau be $$
\hat\tau^{(k)}_{j_1, j_2}=\frac{2}{n(n-1)}\sum_{i,i'\in[n]}\text{sign}\{(X_{ij_1}^{(k)}-X_{i'j_1}^{(k)})(X_{ij_2}^{(k)}-X_{i'j_2}^{(k)})\},
$$
and then estimate the correlation matrices $R^{(k)}=(R^{(k)}_{j_1,j_2})_{j_1,j_2\in[p]}$ by $$
\hat R^{(k)}_{j_1,j_2}=\sin(\frac{\pi}{2}\hat\tau^{(k)}_{j_1, j_2})\cdot\1\{j_1\neq j_2\}+1\cdot\1\{j_1= j_2\} \quad\text{ for } k=1,2.
$$
At last, we let $\hat D_V^{(k)}=\diag((\tilde\sigma_{11}^{(k)})^{1/2},..., (\tilde\sigma_{pp}^{(k)})^{1/2})$, and estimate $\Sigma_k$ by $$
\tilde\Sigma_k=\hat D_V^{(k)}\hat R\hat D_V^{(k)}  \text{ for } k=1,2.
$$

Moreover, we estimate the monotone transformation in a pooled way as \begin{align*}
\hat f_j(t)=\frac{1}{n_1+n_2}\Big(&n_1\left(\hat\mu_j^{(1)}+(\hat\sigma_{jj}^{(1)})^{1/2}\cdot\Phi^{-1}\left(\hat F^{(1)}_j(t)\right)\right)\\
&+n_2(\hat\mu_j^{(2)}+(\hat\sigma_{jj}^{(2)})^{1/2}\cdot\Phi^{-1}\left(\hat F^{(2)}_j(t))\right)\Big),
\end{align*} 
where $\tilde\mu_j^{(k)}=\frac{1}{n_k}\sum_{i=1}^{n_k}x^{(k)}_{ij}$, and $\tilde\sigma_{jj}^{(k)}=\frac{1}{n_k-1}\sum_{i=1}^{n_k}(x^{(k)}_{ij}-\tilde\mu_j^{(k)})^2$.

After we obtain the estimators $\tilde\bmu_1,\tilde\bmu_2, \tilde\Sigma_1,  \tilde\Sigma_2, \hat{\bm{f}}$, we can then apply the framework we developed in previous sections for the copula QDA model as follows.

 Firstly, we estimate the $\hat D$ and $\hat\bbeta$ by plugging them into \eqref{opt1} and \eqref{opt2} to get $\tilde D$ and $\tilde\bbeta$ respectively. 

Under the SeDA model, the oracle classification rule is given by 
\begin{equation}\label{e:copula:qda}
G^{\rm copula}_{\bth}(\bm z)  =  
\begin{cases}
1, \quad  (f(\bm z)-\bmu_1)^\top D(f(\bm z)-\bmu_1)-2\bdel^\top\Omega_2(f(\bm z)-\bar \bmu)-\log({|\Sigma_1|\over |\Sigma_2|})+2\log(\frac{\pi_1}{\pi_2}) >0\\
2, \quad  (f(\bm z)-\bmu_1)^\top D(f(\bm z)-\bmu_1)-2\bdel^\top\Omega_2(f(\bm z)-\bar \bmu)-\log({|\Sigma_1|\over |\Sigma_2|})+2\log(\frac{\pi_1}{\pi_2}) \le0,
\end{cases} 
\end{equation}
Therefore, for a new  observation $\bm z$, we propose the following extended classification rule Copula SDAR (CSDAR) for the QDA under the copula model.
\begin{align}\label{SDAR-copula}
&\hat G_{\rm CSDAR}(\bm z)=1+\\
\notag&{\1}\{(\hat{\bm f}(\bm z)-\tilde\bmu_1)^\top \hat D(\hat{\bm f}(\bm z)-\tilde\bmu_1)-2\tilde\bbeta^\top(\hat{\bm f}(\bm z)-\frac{\tilde\bmu_1+\tilde\bmu_2}{2})-\log(|\tilde D\tilde\Sigma_1+I_p|)+\log(\frac{\hat\pi_1}{\hat\pi_2}) \le 0\}.
\end{align}
%\begin{equation}
%\hat D=\arg\min_{D\in\R^{p\times p}} \left\{|D|_1: \, |\frac{1}{2}\hat\Sigma_1 D \hat\Sigma_2+\frac{1}{2}\hat\Sigma_2 D \hat\Sigma_1-\hat\Sigma_1+\hat\Sigma_2|_{\infty}\le\lambda_{1,n}\right\}
%\label{opt1.c},
%\end{equation}
%
%\begin{equation}
%\hat\bbeta=\arg\min_{\bbeta\in\R^{p}}\left\{\|\bbeta\|_1: \, \| \hat\Sigma_2 \bbeta -\hat\bmu_2+\hat\bmu_1\|_{\infty}\le\lambda_{2,n}\right\}
%\label{opt2.c},
%\end{equation}
%
%For a new target observation $ x^*=(x_1^*, ..., x_p^*)$, and let the prediction \begin{equation}\label{pred}
%\hat \mu^*=\hat f_0^{-1}(\sum_{i=1}^p\hat f_i(x_i^*)\hat\beta(\lambda)_i ),
%\end{equation}
%where $\hat f_0^{-1}$ is the generalized inverse of $\hat f_0$:$$
%\hat f_0^{-1}(t)=\inf\{x\in\R: \hat f_0(x)\ge t\}.
%$$
%
%Then $\hat \mu^*$ is consistent, as the result established below.
We then derive the theoretical properties for this extended SDA rule. 
At first, we have the following bounds on estimating $\bbeta$ and $D$ in this non-Gaussian setting.
\begin{theorem}\label{para-copula}
Consider the parameter space $\Theta_p(s_1, s_2)$, and assume that $n_1\asymp n_2$, $(s_1+s_2)\cdot n^{-1}\to 0$ depending on $\{ F_i(z_i)\}_{i=1}^p$. Then the proposed SDAR classification rule in \eqref{SDAR} satisfies that, with probability at least $1-O(p^{-1})$,
\begin{align*}
&\|\tilde D-D\|_F\lesssim \sqrt{\frac{s_1\log p}{n^{}}}; \quad \|\tilde\bbeta-\bbeta\|_2\lesssim \sqrt{\frac{s_2\log p}{n^{}}}.
%&\|\hat D-D\|_F\lesssim \sqrt{\frac{s_1\log p}{n}};\\
%&\|\hat\bbeta-\bbeta\|_2\lesssim \sqrt{\frac{s_2\log p}{n}}.
\end{align*}
\end{theorem}
 
 We then analyze the misclassification error of this extended SDAR rule. In addition to parameter estimation, under the SeDA model, it's required to estimate $\{f_j(t)\}_{j=1}^p$. Since the estimation of $f_j$ is only accurate when there are sufficient samples around $t$, we define the following region: let $S={j_1,...,j_s}$ be the joint set of the row support of $D$ and support of $\bbeta$, and $\gamma\in(0,1)$, define $M_n\subset \R^p$:
 \begin{align*}
 M_n=\{\bm x\in\R^p: \bm x_S\in&[f_{j_1}^{-1}(-\sqrt{2\gamma\log n}),f_{j_1}^{-1}(\sqrt{2\gamma\log n})]\times\ldots\\
 &\times[f_{j_s}^{-1}(-\sqrt{2\gamma\log n}),f_{j_s}^{-1}(\sqrt{2\gamma\log n})]\},
 \end{align*}
 which is a high-probability event with $\Pro(M_n)\ge 1- C\cdot s\cdot n^{-\gamma}$.
 
We then define the misclassification error for the copula model. 
\begin{equation}\label{risk-copula}
 \tilde R_{\bth}(\hat G)=\E_{\bth}[ \1\{\hat G(\bm z)\neq L(\bm z)  \mid \bm z\in M_n\}].
\end{equation}

Similar construction of $M_n$ has been considered in all previous papers considering the SeDA model \citep{han2013coda,zhao2014semiparametric,mai2015sparse}. We then have the following result for the misclassification error.

\begin{theorem}\label{Rn-copula}
Consider the parameter space $\Theta_p(s_1, s_2)$. Under the same condition as in Theorem~\ref{para-copula}, and $\gamma\in(0,1)$ satisfies $s\cdot n^{-\gamma}\to 0$, then the proposed SDAR classification rule in \eqref{SDAR} satisfies that, for sufficiently large $n$,
$$\sup_{\bth\in\Theta_p(s_1, s_2)}\E\left[\tilde R_{\bth}(\hat G_{\rm CSDAR})-\tilde R_{\bth}(G^{\rm copula}_{\bth})\right]\lesssim  (s_1+{s_2})\cdot {\frac{\log p}{n^{1-\gamma}}}\cdot\log^2 n.$$
\end{theorem}

\begin{remark} 
The additional term $n^\gamma$ commonly appeared in recent literature studying Gaussian copula models, especially for the classification setting, see \citet{han2013coda, zhao2014semiparametric} and \citet{mai2015sparse}, and this term occurs due to the necessity of estimating $f_j's$. We improve the the convergence rate of $\frac{\sqrt s}{n^{(1-\gamma)/2}}$ from prior works to $\frac{s}{n^{1-\gamma}}$ in Theorem \ref{Rn-copula}.  
%The condition $\max_{i=1, ...,p}  F_i(z_i)\in(\delta_c, 1-\delta_c)$ for some $\delta_c>0$ can be removed, but then the convergence rate would become, for any $\gamma>0$, if $(s_1+s_2)\cdot n^{-c\gamma}\to 0$ for some constant $c>0$ depending on $\{ F_i(z_i)\}_{i=1}^p$,
%$$\sup_{\bth\in\Theta_p(s_1, s_2)}\E\left[R_{\bth}(\hat G_{\rm SDAR-ext})-R_{\bth}(G^*_{\bth})\right]\lesssim  (s_1+{s_2})\cdot {\frac{\log p}{n^{1-\gamma}}}\cdot\log^2 n.$$
\end{remark}

%%%%%%%%%%%%%%%%%%%%%%%%%%%%%%%%%%%%%%%%%%%
%%%%%%%%%%%%%%%%%%%%%%%%%%%%%%%%%%%%%%
\section{Numerical Studies}
\label{sec:simulations}
%%%%%%%%%%%%%%%%%%%%%%%%%%%%%%%%%%%%%%

In this section we firstly conduct simulation studies to investigate the impossibility results shown in Section \ref{sec:general}.2, and then study numerical properties of the proposed SDAR and CSDAR methods under various settings.
%%%%%%%%%%%%%%%%%%%%%%%%%%%%%
\subsection{Impossibility results}
%%%%%%%%%%%%%%%%%%%%%%%%%%%%%

We would like to illustrate the impossibility results Theorem \ref{general-lb1} and Theorem \ref{general-lb2} in a numerical fashion in this subsection.

Let us start with Theorem \ref{general-lb1}, which shows the sparsity condition on $\bbeta$ is necessary. In the simulation, we consider the simple case where both covariance matrices are known to be identity but the means are unknown: $\bm x_1,...,\bm x_{n} \sim N_p(\bmu_1,I_p)$ and  $\bm y_1, ..., \bm y_{n}\sim N_p(\bmu_2, I_p)$ and let $\bmu_1=-\bmu_2=\bmu=\frac{1}{\sqrt p}\cdot\bm 1_p$, satisfying $\|\bmu_1-\bmu_2\|_2=2$. 

We consider nine cases where $(n,p)=(100, 200)$, $(150,200)$, $(200,200)$, $(100,300)$, $(200,300)$, $(300,300), (200, 600), (400,600), (600,600)$. In each setting, we compare the oracle classification rule $G^*_{\bth}$ in \eqref{e:intro:qda} with the plug-in classification rule $\hat G$ where we estimate $\bmu_1, \bmu_2$ by the sample means. The testing sample size is set to $100$ and the simulation is repeated $100$ times in each setting. The simulations results is summarized in the following table. 
 
\begin{table}[H]
\centering
\caption{Average classification errors (s.e.) based on $n = 100$ test samples from 100 replications under the setting where covariance matrices are known to be identity.}
\label{label1}
\begin{tabular}{l|c|c|c}
      & $n$   & $R_{\bth}(\hat G)$        & $R_{\bth}(G_{opt})$        \\ \hline
      & 100 & 0.242 (0.054) & 0.155 (0.035) \\
p=200 & 150 & 0.232 (0.051) & 0.155 (0.035) \\
      & 200 & 0.219 (0.039) & 0.155 (0.035) \\ \hline
      & 100 & 0.265 (0.048) & 0.149(0.032)  \\
p=300 & 200 & 0.223 (0.047) & 0.149(0.032)  \\
      & 300 & 0.208 (0.038) & 0.149(0.032)  \\ \hline
      & 200 & 0.269 (0.045) & 0.158 (0.035) \\
p=600 & 400 & 0.230 (0.035) & 0.158 (0.035) \\
      & 600 & 0.201 (0.035) & 0.158 (0.035)
\end{tabular}
\end{table}

To illustrate Theorem \ref{general-lb2}, we consider a simple case where $\bmu_1=-\bmu_2=(1,0,0,...,0)^\top$ and the covariance matrices are known to be diagonal.  Two classes are $N_p(\bmu_1, I_p)$ and $N_p(\bmu_2, \Sigma_2)$, where $\Sigma_2=(I_p+\sum_{i=1}^{p/2}\frac{2}{\sqrt p}E_{i,i})^{-1}$ and $E_{i,i}$ is a $p\times p$ matrix whose $(i,i)$-th entry is $1$ and $0$ else.

 We consider nine cases where $(n,p)=(100, 200)$, $(150,200)$, $(200,200)$, $(100,300)$, $(200,300)$, $(300,300), (200, 600), (400,600), (600,600)$. In each setting, we compare the oracle classification rule $G_{opt}$, that is \eqref{e:intro:qda}, with the plug-in classification rule $\hat G$ where we estimate $\Sigma_1, \Sigma_2$ by the diagonals of sample covariance matrices.  The following table summarizes the simulation results where the testing sample size is set to $100$ and the simulation is repeated $100$ times.

\begin{table}[H]
\centering
\caption{Average classification errors (s.e.) based on $n = 100$ test samples from 100 replications under the setting where means are known to be $\bm 0_p$ and covariance matrices are known to be diagonal.}
\label{label2}
\begin{tabular}{l|c|c|c}
      & $n$   & $R_{\bth}(\hat G)$        & $R_{\bth}(G_{opt})$    \\ \hline
      & 100 & 0.274 (0.049) & 0.193 (0.038) \\
p=200 & 150 & 0.260 (0.036) & 0.193 (0.038) \\
      & 200 & 0.252 (0.033) & 0.193 (0.038) \\ \hline
      & 100 & 0.271 (0.043) & 0.151(0.034)  \\
p=300 & 200 & 0.238 (0.048) & 0.151(0.034)  \\
      & 300 & 0.224 (0.039) & 0.151(0.034)  \\ \hline
      & 200 & 0.296 (0.032) & 0.183 (0.046) \\
p=600 & 400 & 0.255 (0.055) & 0.183 (0.046) \\
      & 600 & 0.245 (0.037) & 0.183 (0.046)
\end{tabular}
\end{table}

%%%%%%%%%%%%%%%%%%%%%%%%%%%%%
\subsection{SDAR on synthetic data}
%%%%%%%%%%%%%%%%%%%%%%%%%%%%%

In this section, we provide extensive numerical evidence to show the empirical performance of SDAR by comparing it to its competitors, including the sparse QDA (SQDA, Li and Shao (2015)), %the innovated interaction screening for sparse quadratic discriminant analysis (IIS- SDAR, Fan et al. (2015b)), 
the direct approach for sparse LDA (LPD, Cai and Liu  (2012)), the conventional LDA (LDA), the conventional QDA (QDA) and the oracle procedure (Oracle). The oracle procedure uses the true underlying model and serves as the optimal risk bound for comparison. We also compare SDAR with model-free classifiers, including random forest (RF), AdaBoost (AB), SVM, and Kernel SVM (KSVM). We evaluate all methods via three synthetic datasets.% and two real datasets.

In all simulations, the sample size is $n_1=n_2 = 200$ while the number of variables $p$ varies from $100, 200, 400$ to $600$. The sparsity levels are set to be $s_1=10, s_2=20$. The discriminating direction $\bbeta = (1,\ldots,1,0,\ldots,0)^\top$ is sparse such that only the first $s_1 = 10$ entries are nonzero. Given the inverse covariance matrix of the second sample $\Omega_2$, the mean for class 1 is $\bmu_1 = (0,\ldots, 0)^\top$ and  the mean for class 2 is set to be $\bmu_2=\bmu_1 - \Sigma_2 \bbeta $. In addition, the differential graph $D$ is a random sparse symmetric matrix with its nonzero positions generated by uniform sample. Each nonzero entry on $D$ is $i.i.d.$ and from a standard normal distribution $N(0,1)$. Lastly, we let $\Omega_1=D+\Omega_2$, and $\Omega_1=\Sigma_1^{-1},\Omega_2=\Sigma_2^{-1}$. We use the following three models to generate $\Omega_2$.

\begin{description}

\item[Model 1:] {\bf Block sparse model:}  We generate $\Omega_2=U^T \Lambda U$, where $\Lambda \in \R^{p\times p}$ is a diagonal matrix and its entries are $i.i.d.$ and uniform on $[1, 2]$, and $U\in \R^{p\times p} $ is a random matrix with $i.i.d.$ entries from $N (0, 1)$.  In the simulation,  the tuning parameters for SDAR method are chosen over a grid $\{\frac{k}{2}\sqrt{\frac{\log p}{n}}\}_{k=1:15}$.

\item[Model 2:] {\bf AR(1) model:} $\Omega_2 =(\Omega_{ij})_{p\times p}$ with $\Omega_{ij} = \rho^{|i-j|}$.  In the simulation,  the tuning parameters for the SDAR method are chosen by cross validation over a grid $\{\frac{k}{4}\sqrt{\frac{\log p}{n}}\}_{k=1:15}$. The simulation results from $100$ replications are summerized as follows, with $\rho=0.5$.

\item[Model 3:] {\bf Erd\H{o}s-R\'{e}nyi random graph:} Let $\tilde \Omega_2 = (\tilde \omega_{ij})$ where $\tilde \omega_{ij} = u_{ij} \delta_{ij}$, $\delta_{ij}\sim {\rm Ber}(1,\rho)$ being the Bernoulli random variable with success probability 0.05 and $u_{ij}\sim {\rm Unif}[0.5,1]\cup[-1, -0.5]$. After symmetrizing $\tilde \Omega_2$, set $\Omega_2 = \tilde \Omega_2 + \{\max(- \phi_{\min}(\tilde \Omega_2), 0) + 0.05 \} {\bf I}_p$ to ensure the positive definiteness.   In the simulation,  the tuning parameters for SDAR method are chosen over a grid $\{\frac{k}{2}\sqrt{\frac{\log p}{n}}\}_{k=1:15}$. 

\end{description}

In each model, the number of repetition is set to be 100, and the classification errors are evaluated based on the test data with size $200$ that is generated from a Gaussian mixture model $\frac{1}{2} N_p(\bmu_1, \Sigma_1)+\frac{1}{2} N_p(\bmu_2, \Sigma_2)$. We compare the proposed SDAR method with the oracle QDA rule \eqref{e:intro:qda}. The simulation results are summarized in Table \ref{table1}.

This simulation results show that the proposed SDAR algorithm outperforms the LPD algorithm when there are strong interactions among features ($D\neq 0$). As expected, the conventional LDA and QDA works poorly in the high-dimensional setting, and the performance of conventional QDA is even worse due to overfitting. Comparing to the model-free classifiers, we found that they perform better than LDA/QDA, but still have higher misclassification error than the proposed SDAR algorithm since the latter incorporates more model information such as Gaussianity and sparsity. In the setting where $D=0$, the estimated $\hat D$ would equal to $D=0$ for properly chosen $\lambda_1$, according to Theorem~\ref{estimation}. As we estimate $\bbeta$ and $D$ separately, the proposed SDAR rule in this case would adaptively reduced to LPD. For reasons of space we do not present the detailed numerical results for this case.

\subsection{CSDAR on synthetic data under Gaussian copula model}
Same as the previous discussion, in this section, we compare the performance of CSDAR with its competitors, including LDA, QDA, SQDA, LPD, RF, AB, SVM and KSVM. For the synthetic data generation, we use the same parameter settings as Model 1- Model 3 to generate mean vectors and covariance matrices, and call them Model 4-6. Additionally, after the generation of Gaussian distributed data, for each model, we apply the following monotone transformations: $f_1(x)=x^3$, $f_2(x)=\arctan(x)$, $f_3(x)=\arctan^3(x)$, $f_4(x)=x^5$ to the $1^{st}-5^{th}$, $11^{th}-15^{th}$, $21^{st}-50^{th}$, and $51^{st}-85^{th}$ entries respectively. The simulation results are summarized in Table \ref{table2}.

This simulation results show that all Gaussian-model based algorithms fail in this setting, while the proposed CSDAR classifier and model free algorithms still maintain their good performances. Further, due to the incorporation of the model information such as Gaussian copula and sparsity, the CSDAR algorithm has smaller misclassification errors than RF, AB, SVM and KSVM in most cases. 
%xtest(:,1:5)=xtest(:,1:5).^3;
%ytest(:,1:5)=ytest(:,1:5).^3;
%xtest(:,11:15)=atan(xtest(:,11:15));
%ytest(:,11:15)=atan(ytest(:,11:15));
%xtest(:,21:50)=atan(xtest(:,21:50)).^3;
%ytest(:,21:50)=atan(ytest(:,21:50)).^3;
%xtest(:,51:85)=xtest(:,51:85).^5;
%ytest(:,51:85)=ytest(:,51:85).^5;
%ztest = [xtest;ytest];

%\begin{table}[H]
%\centering
%\caption{Average classification errors (s.e.) based on $n = 100$ test samples from 100 replications under three different models}
%\label{table1}
%\begin{tabular}{|l|ll|ll|ll|}
%\hline
%        & p=60                               &               & p=80                               &               & p=100                              &               \\ \hline
%        & \multicolumn{1}{l|}{SDAR}          & Oracle        & \multicolumn{1}{l|}{SDAR}          & Oracle        & \multicolumn{1}{l|}{SDAR}          & Oracle        \\ \hline
%Model 1 & \multicolumn{1}{l|}{0.045(0.014)}  & 0.040(0.012) & \multicolumn{1}{l|}{0.039(0.012)} & 0.034(0.013) & \multicolumn{1}{l|}{0.045(0.013)}         & 0.036(0.011)        \\ \hline
%Model 2 & \multicolumn{1}{l|}{0.045(0.013)} & 0.038(0.014)  & \multicolumn{1}{l|}{0.044(0.015)} & 0.038(0.013) & \multicolumn{1}{l|}{0.047(0.013)}  & 0.041(0.012) \\ \hline
%Model 3 & \multicolumn{1}{l|}{0.121(0.069)} & 0.108(0.035) & \multicolumn{1}{l|}{0.143(0.058)} & 0.126(0.032) & \multicolumn{1}{l|}{0.151(0.057)} & 0.140(0.026) \\ \hline
%\end{tabular}
%\end{table}

\begin{table}[t]
\caption{Average classification errors (s.d.) based on $n = 200$ test samples from 100 replications under three different models (Gaussian setting)}
\begin{tabular}{lccccc}\label{table1}
        & $p$                & 100          & 200          & 400            & 600          \\ \hline
        & LDA                & 0.200(0.019) & 0.224(0.028) & 0.269(0.022)   & 0.302(0.024) \\
        & QDA                & 0.236(0.026) & 0.274(0.023) & 0.418(0.025)   & 0.432(0.027) \\
        & SQDA (Shao et al.) & 0.202(0.022) & 0.231(0.027) & 0.301(0.023)   & 0.347(0.025) \\
        & LPD                & 0.151(0.020) & 0.163(0.021) & 0.208(0.028)   & 0.256(0.025) \\
Model 1 & RF                 & 0.176(0.021) & 0.019(0.022) & 0.225(0.018)   & 0.231(0.013) \\
        & Adaboost           & 0.182(0.018) & 0.210(0.029) & 0.229(0.017)   & 0.225(0.026) \\
        & SVM                & 0.467(0.062) & 0.453(0.076) & 0.415(0.049)   & 0.447(0.061) \\
        & KSVM               & 0.213(0.027) & 0.254(0.016) & 0.279(0.022)   & 0.259(0.029) \\
        & SDAR               & {\bf 0.117(0.019)} & {\bf 0.159(0.022)} & {\bf 0.191(0.029)} & {\bf 0.202(0.027)} \\
        & CSDAR               & { 0.132(0.017)} & {0.173 (0.025)} & {0.209(0.024)} & {0.217(0.022)} \\
        & Oracle             & 0.076(0.010) & 0.097(0.007) & 0.098(0.010)   & 0.097(0.009) \\ \hline
        & LDA                & 0.231(0.022) & 0.214(0.021) & 0.335(0.025)   & 0.378(0.027) \\
        & QDA                & 0.249(0.025) & 0.296(0.029) & 0.405(0.026)   & 0.446(0.028) \\
        & SQDA (Shao et al.) & 0.214(0.023) & 0.243(0.024) & 0.327(0.023)   & 0.376(0.025) \\
        & LPD                & 0.163(0.018) & 0.156(0.019) & 0.220(0.027)   & 0.253(0.024) \\
Model 2 & RF                 & 0.199(0.027) & 0.272(0.020) & 0.339(0.038)   & 0.370(0.029) \\
        & Adaboost           & 0.200(0.018) & 0.229(0.017) & 0.268(0.027)   & 0.279(0.031) \\
        & SVM                & 0.467(0.059) & 0.481(0.040) & 0.474(0.045)   & 0.489(0.026) \\
        & KSVM               & 0.215(0.031) & 0.304(0.021) & 0.331(0.022)   & 0.336(0.018) \\
        & SDAR               & {\bf 0.141(0.015)} & {\bf 0.152(0.019)} & {\bf 0.155(0.020)}   & {\bf 0.192(0.019)} \\
        & CSDAR               & { 0.159(0.021)} & {0.163 (0.019)} & {0.183(0.026)} & {0.233(0.027)} \\
        & Oracle             & 0.045(0.010) & 0.054(0.007) & 0.042(0.008)   & 0.056(0.008) \\ \hline
        & LDA                & 0.279(0.028) & 0.305(0.032) & 0.340(0.031)   & 0.387(0.029) \\
        & QDA                & 0.298(0.024) & 0.356(0.025) & 0.406(0.026)   & 0.457(0.025) \\
        & SQDA (Shao et al.) & 0.242(0.024) & 0.294(0.029) & 0.335(0.026)   & 0.374(0.026) \\
        & LPD                & 0.236(0.023) & 0.205(0.020) & 0.234(0.031)   & 0.252(0.027) \\
Model 3 & RF                 & 0.288(0.014) & 0.317(0.022) & 0.343(0.024)   & 0.359(0.027) \\
        & Adaboost           & 0.275(0.028) & 0.272(0.016) & 0.276(0.018)   & 0.252(0.019) \\
        & SVM                & 0.495(0.08)  & 0.477(0.037) & 0.467(0.037)   & 0.461(0.039) \\
        & KSVM               & 0.271(0.034) & 0.325(0.025) & 0.325(0.036)   & 0.313(0.027) \\
        & SDAR               & {\bf 0.115(0.022)} & {\bf 0.137(0.026)} & {\bf 0.146(0.028)}  & {\bf 0.155(0.026)} \\
        & CSDAR               & { 0.143(0.019)} & {0.184 (0.024)} & {0.202(0.032)} & {0.178(0.023)} \\
        & Oracle             & 0.065(0.013) & 0.039(0.009) & 0.031(0.008)   & 0.048(0.010)
\end{tabular}
\end{table}

\begin{table}[t]
\caption{Average classification errors (s.d.) based on $n = 200$ test samples from 100 replications under three different models (Gaussian copula setting)}
\begin{tabular}{lccccc}\label{table2}
        & $p$                & 100                                       & 200                                       & 400                                       & 600                                       \\ \hline
        & LDA                & 0.369(0.011)                              & 0.362(0.013)                              & 0.411(0.023)                              & 0.382(0.014)                              \\
        & QDA                & 0.332(0.012)                              & 0.382(0.008)                              & 0.446(0.009)                              & 0.497(0.002)                              \\
        & SQDA (Shao et al.) & 0.401(0.019)                              & 0.374(0.034)                              & 0.315(0.032)                              & 0.345(0.027)                              \\
        & LPD                & 0.424(0.026)                              & 0.335(0.045)                              & 0.292(0.041)                              & 0.298(0.008)                              \\
Model 4 & RF                 & 0.138(0.005)                              & 0.200(0.008)                              & 0.230(0.011)                              & 0.267(0.023)                              \\
        & Adaboost           & 0.145(0.004)                              & 0.201(0.005)                              & 0.219(0.012)                              & 0.232(0.006)                              \\
        & SVM                & 0.482(0.004)                              & 0.483(0.012)                              & 0.499(0.001)                              & 0.494(0.005)                              \\
        & KSVM               & 0.209(0.008)                              & 0.265(0.010)                              & 0.267(0.007)                              & 0.309(0.008)                              \\
        & CSDAR              & \multicolumn{1}{l}{\textbf{0.125(0.005)}} & \multicolumn{1}{l}{\textbf{0.164(0.005)}} & \multicolumn{1}{l}{\textbf{0.196(0.002)}} & \multicolumn{1}{l}{\textbf{0.206(0.005)}} \\ \hline
        & LDA                & 0.431(0.008)                              & 0.398(0.007)                              & 0.462(0.008)                              & 0.440(0.011)                              \\
        & QDA                & 0.421(0.008)                              & 0.379(0.008)                              & 0.439(0.009)                              & 0.499(0.001)                              \\
        & SQDA (Shao et al.) & 0.455(0.011)                              & 0.392(0.013)                              & 0.388(0.017)                              & 0.417(0.020)                              \\
        & LPD                & 0.451(0.020)                              & 0.428(0.024)                              & 0.405(0.020)                              & 0.431(0.024)                              \\
Model 5 & RF                 & 0.213(0.007)                              & 0.248(0.007)                              & 0.331(0.011)                              & 0.342(0.006)                              \\
        & Adaboost           & 0.203(0.007)                              & 0.225(0.005)                              & 0.246(0.006)                              & 0.265(0.009)                              \\
        & SVM                & 0.489(0.003)                              & 0.485(0.010)                              & 0.491(0.006)                              & 0.499(0.001)                              \\
        & KSVM               & 0.254(0.006)                              & 0.307(0.007)                              & 0.343(0.008)                              & 0.349(0.007)                              \\
        & CSDAR              & \multicolumn{1}{l}{\textbf{0.157(0.018)}} & \multicolumn{1}{l}{\textbf{0.162(0.005)}} & \multicolumn{1}{l}{\textbf{0.160(0.006)}} & \multicolumn{1}{l}{\textbf{0.197(0.009)}} \\ \hline
        & LDA                & 0.351(0.016)                              & 0.403(0.009)                              & 0.439(0.010)                              & 0.406(0.007)                              \\
        & QDA                & 0.416(0.006)                              & 0.426(0.011)                              & 0.435(0.007)                              & 0.489(0.003)                              \\
        & SQDA (Shao et al.) & 0.313(0.016)                              & 0.400(0.022)                              & 0.459(0.022)                              & 0.452(0.012)                              \\
        & LPD                & 0.290(0.013)                              & 0.396(0.018)                              & 0.464(0.014)                              & 0.429(0.013)                              \\
Model 6 & RF                 & 0.294(0.007)                              & 0.290(0.008)                              & 0.335(0.006)                              & 0.266(0.010)                              \\
        & Adaboost           & 0.282(0.008)                              & 0.249(0.007)                              & 0.263(0.009)                              & 0.221(0.015)                              \\
        & SVM                & 0.452(0.007)                              & 0.493(0.007)                              & 0.494(0.007)                              & 0.499(0.001)                              \\
        & KSVM               & 0.304(0.007)                              & 0.312(0.008)                              & 0.370(0.012)                              & 0.330(0.014)                              \\
        & CSDAR              & \textbf{0.209(0.008)}                     & \textbf{0.189(0.004)}                     & \textbf{0.172(0.007)}                     & \textbf{0.165(0.005)}                    
\end{tabular}
\end{table}

\subsection{Real data}
\label{sec:real:data}
%%%%%%%%%%%%%%%%%%%%%%%%%%%%%%%%%%%%%%

In addition to the simulation studies, we also illustrate the merits of the SDAR classifier in the analysis of two real datasets to further investigate the numerical performance of the proposed method. One is the prostate cancer data in Singh, et al. (2002), which is available at \url{ftp://stat.ethz.ch/Manuscripts/dettling/prostate.rda}, and another dataset is the colon tissues data analyzed in Alon et al. (1999) by using the Oligonucleotide microarray technique, available at \url{ http://microarray.princeton.edu/oncology/affydata/index.html}. These two datasets were frequently used for illustrating the empirical performance of the classifier  for high-dimensional data in recent literature, see Dettling (2004) and Efron (2010). We will compare SDAR with the existing methods, including the sparse QDA (SQDA, Li and Shao (2015)), the direct approach for sparse LDA (LPD, Cai and Liu  (2012)), the conventional LDA (LDA), the conventional QDA (QDA). 

\subsubsection{Prostate cancer data}
The prostate cancer data consists of genetic expression levels for $p = 6033$ genes from 102 individuals (50 normal control subjects and 52 prostate cancer patients). The SDAR classifier allows us to model the interactions among genes and thus improve the classification accuracy. For this data, we follow the same data cleaning routine in Cai and Liu (2011), retaining only the top 200 genes with the largest absolute values of the two sample $t$-statistics. The average
classification errors using 5-fold cross-validation for various methods with 50 repetitions are reported in Table~\ref{prostate}. The proposed SDAR method outperforms all the other methods
\begin{table}[H]
\centering
\caption{Classification error(\%) with s.d. of prostate cancer data by various methods}
\label{prostate}
\begin{tabular}{|ccclll|}
\hline
                                          & SDAR & SQDA (Shao et al.) & LPD & LDA   & QDA     \\ \hline
\multicolumn{1}{|c}{Testing error} &    2.20 (1.11)      &     3.10 (1.26)     & 11.20 (1.87) & 32.20 (3.67) & 35.30 (4.18)      \\ \hline
\multicolumn{1}{|c}{ } &    CSDAR     &     RF     & AB & SVM & KSVM      \\ \hline
\multicolumn{1}{|c}{Testing error} &    4.27 (0.15)      &     3.10 (4.26)     & 5.36 (4.89) & 42.20 (3.21) & 5.36 (2.39)      \\ \hline
\end{tabular}
\end{table}
\subsubsection{Colon tissues data}
The colon tissues data analyzed gene expression difference between tumor and normal colon tissues using the Oligonucleotide microarray technique, consisting 20 observations from normal tissues and 42 observations
from tumor tissues, measured in $p=2000$ genes.

Similarly to the analysis of the prostate cancer data, to control the computational costs, we use 200 genes with the largest absolute values of the two sample $t$-statistics. Classification results by using  5-fold cross-validation with 50 repetitions are summarized in Table \ref{colon}. In this example, the SDAR is still the best among all classifiers. 
\begin{table}[H]
\centering
\caption{Classification error(\%) with s.d. of colon tissues data by various methods}
\label{colon}
\begin{tabular}{|ccclll|}
\hline
                                          & SDAR & SQDA (Shao et al.) & LPD & LDA   & QDA     \\ \hline
\multicolumn{1}{|c}{Testing error} &    19.05 (2.40)      &     23.20 (2.36)     & 26.67 (2.75)  & 38.20 (3.14) & 39.30 (4.71)       \\ \hline
\multicolumn{1}{|c}{ } &    CSDAR     &     RF     & AB & SVM & KSVM      \\ \hline
\multicolumn{1}{|c}{Testing error} &    22.27 (2.41)      &     25.33 (4.24)     & 23.81 (3.72) & 46.20 (0.32) & 25.71 (4.24)      \\ \hline
\end{tabular}
\end{table}

\section{Extension to the Multi-group Classification}
\label{sec:extension}
%%%%%%%%%%%%%%%%%%%%%%%%%%%%%%%%%%%%%%

We have so far focused on high-dimensional QDA for two groups in the Gaussian setting. The methodology and theory developed in the earlier sections can be extended to multi-group classification and to classification under the Gaussian copula model. 

%%%%%%%%%%%%%%%%%%%%%%%%%%%%%%%%%%%%%%
\subsection{Multi-group classification}
%%%%%%%%%%%%%%%%%%%%%%%%%%%%%%%%%%%%%%

We first turn to  multi-group classification. Suppose there are $K$ classes $N_p(\bmu_k, \Sigma_k)$ with prior probability $\pi_k$ for $1 \le k \le K$ respectively, and an observation $\bm z$ is drawn from the same distribution. In the ideal setting where all the parameters are known, the oracle rule classifies $\bm z$ to class $k$ if and
only if
$$
k=\arg\min_{k\in[K]}\left\{Q_k(\bm z) \right\},
$$
%{\red Double check. The first and last terms are both $\log\pi_k$.}
where the discriminating function $Q_k(\bm z)$ is $$
Q_k(\bm z) =  
\begin{cases}
1, \quad  k=1\\
\frac{1}{2}(\bm z-\bmu_k)^\top D_k(\bm z-\bmu_k)-\bbeta_k^\top(\bm z-\bar\bmu_k)-\frac{1}{2}\log| D_k\Sigma_1+I_p|+\log\pi_k, \quad  k\ge 2,
\end{cases} 
$$
with $\bar\bmu_k=\frac{\bmu_1+\bmu_k}{2}, D_k=\Omega_1-\Omega_k$, $\bbeta_k=\Omega_1(\bmu_k-\bmu_1)$, and $\Omega_k=\Sigma_k^{-1}$.
When the parameters are unknown and random samples from $K$ classes (with prior probabilities $\{\pi_k\}_{k=1}^K$) are available: $\bm x^{(k)}_1,...,\bm x^{(k)}_{n_k} \stackrel{i.i.d.}{\sim} N_p(\bmu_k, \Sigma_k)$, $k=1, ..., K$,  by assuming the sparsity on $D_k$'s and $\bbeta_k$'s, they can then be estimated by solving a similar linear programming as in \eqref{opt1} and \eqref{opt2}. For $k=2,3,...,K$, $D_k$ and $\bbeta_k$ are estimated by 
\begin{equation}
\hat D_k=\arg\min_{D\in\R^{p\times p}} \left\{|D|_1: \, |\frac{1}{2}\hat\Sigma_1 D \hat\Sigma_k+\frac{1}{2}\hat\Sigma_2 D \hat\Sigma_1-\hat\Sigma_1+\hat\Sigma_k|_{\infty}\le\lambda_{1,n}\right\}
\label{optk},
\end{equation}
where $\lambda_{1,n}$ is a tuning parameter with constant $c_1>0$. 
\begin{equation}
\hat\bbeta_k=\arg\min_{\bbeta\in\R^{p}}\left\{\|\bbeta\|_1:\, \| \hat\Sigma_1 \bbeta -\hat\bmu_k+\hat\bmu_1\|_{\infty}\le\lambda_{2,n}\right\}
\label{optk2},
\end{equation}
where $\lambda_{2,n}$ is a tuning parameter with constant $c_2>0$.

 Given these estimators and $\hat\pi_k=n_k/(\sum_{k=1}^K n_k)$, the discriminating function is then estimated by$$
\hat Q_k(\bm z) =  
\begin{cases}
1, \quad  k=1\\
\frac{1}{2}(\bm z-\hat\bmu_k)^\top \hat D_k(\bm z-\hat\bmu_k)-\hat\bbeta_k^\top(\bm z-\hat{\bar\bmu}_k)-\frac{1}{2}\log| \hat D_k\hat\Sigma_1+I_p|+\log\hat\pi_k, \quad  k\ge 2,
\end{cases} 
$$
Then the SDAR classification rule for multi-group classification is constructed  as 
$$
\hat G(\bm z)=\argmin_{k\in[K]}\{\hat Q_k(\bm z)\}.
$$

 By applying the same techniques we developed for Theorems \ref{estimation} and \ref{Rn}, similar convergence rates can be obtained for both estimation and classification errors.

%%%%%%%%%%%%%%%%%%%%%%%%%%%%%%%%%%%%%%
\section{Proofs}
\label{sec:proof}
%%%%%%%%%%%%%%%%%%%%%%%%%%%%%%%%%%%%%%
We present the proofs of Theorems \ref{general-lb1}, \ref{general-lb2}, \ref{estimation}, \ref{Rn} in this section. The proof of Theorem~\ref{sparse-lb} is similar to Theorems \ref{general-lb1}, \ref{general-lb2}, so we present its proof in the supplement.% 

%%%%%%%%%%%%%%%%%%%%%%%%%%%%%
\subsection{Proof of Theorem \ref{general-lb1} and \ref{general-lb2}}\label{sec:pf:imqda}
%%%%%%%%%%%%%%%%%%%%%%%%%%%%%
 We prove Theorem \ref{general-lb1} and \ref{general-lb2} for the case where $p\lesssim n$. In the case where $\limsup_{n\to\infty} p/n=\infty$, the right hand side of Theorem  \ref{general-lb1} (and \ref{general-lb2})  is of constant order and we can consider only the first $n$-dimension of $p$-dimensional vector, and assume the rest is known. 
 
We begin by collecting a few important technical lemmas that will be used in the proofs of the minimax lower bounds.
%%%%%%%%%%%%%%%%%%%%
\subsubsection{Technical lemmas}
\begin{lemma}[\cite{azizyan2013minimax}]
\label{triangle}
For any $\bth, \tilde \bth \in \Theta_{p}(s_1, s_2)$ and any classification rule $\hat G$, recall that $G^*_{\tilde\bth}$ is the optimal rule w.r.t. $\tilde\bth$. If $$
L_{\bth}(G^*_{\tilde\bth})+L_{\bth}(\hat G)+\sqrt{\frac{KL(\Pro_{\bth}, \Pro_{\tilde \bth})}{2}}\le 1/2,
$$ then$$ 
L_{\bth}(G^*_{\tilde\bth})-L_{\bth}(\hat G)-\sqrt{\frac{KL(\Pro_{\bth}, \Pro_{\tilde \bth})}{2}}\le L_{\tilde \bth}(\hat G)
\le L_{\bth}(G^*_{\tilde\bth})+L_{\bth}(\hat G)+\sqrt{\frac{KL(\Pro_{\bth}, \Pro_{\tilde \bth})}{2}},$$
\end{lemma}
where the KL divergence of  two probability density functions $\Pro_{\bth_1}$ and $\Pro_{\bth_2}$ is defined by $$
KL(\Pro_{\bth_1}, \Pro_{\bth_2})=\int \Pro_{\bth_1}(x)\log\frac{\Pro_{\bth_1}(x)}{\Pro_{\bth_2}(x)}\:dz.
$$

\begin{lemma}[\cite{tsybakov2009introduction}]
\label{fano}
Let $M\ge 0$ and $\bth_0, \bth_1, ... ,\bth_M \in \Theta_p(s_1, s_2)$. For some constants $\alpha\in(0,1/8), \gamma>0$, and any classification rule $\hat G$, if $KL(\Pro_{\bth_i}, \Pro_{\bth_0})\le \alpha \log M/{n}$ for all $1\le i \le  M$, and $L_{\bth_i}(\hat G)<\gamma$ implies $L_{\bth_j}(\hat G)\ge\gamma$ for all $0\le i\neq j\le M$, then$$
\inf_{\hat G} \sup_{i\in[M]}\E_{\bth_i}[L_{\bth_i}(\hat G)]\gtrsim\gamma.
$$
\end{lemma}

To use Fano's type minimax lower bound, we need a covering number argument, provided by the following Lemma \ref{packing}.
\begin{lemma}[\cite{tsybakov2009introduction}]
\label{packing}
Define $\mathcal{A}_{p,s}=\{\bm u:\: \bm u\in\{0,1\}^p, \|\bm u\|_0= s\}$. If $p\ge 4s$, then there exists a subset $\{\bm u_{0},\bm u_{1}, ..., \bm u_{M}\}\subset \mathcal{A}_{p,s}$ such that $\bm u_{0}=\{0,...,0\}^\top$, $\rho_H(\bm u_{i}, \bm u_{j})\ge s/2$ and $\log (M+1)\ge \frac{s}{5}\log(\frac{p}{s})$, where $\rho_H$ denotes the Hamming distance.
\end{lemma}

%\begin{lemma}\label{difference}
%Consider the QDA problem with two distributions $N_p(\bmu_1, \Sigma_1)$ and $N_p(\bmu_2, \Sigma_2)$, and recall that $D=\Omega_2-\Omega_1$. Assume that conditions \textbf{(C1)}, \textbf{(C2)}, \textbf{(C4)} hold, then there exists a universal constant $M>0$, such that $$
%\|D\|_F\le M.
%$$
%\end{lemma}

%%%%%%%%%%%%%%%%%%%%
\subsubsection{Main proof of Theorem \ref{general-lb1}}\label{sec:pfthm1}
%%%%%%%%%%%%%%%%%%%%
At first we construct the following least favorable subset, which characterizes the difficulty of the general QDA problem. Let's consider the parameter space
 \begin{align*}
 \Theta_1=\{&\bth_{\bm u}=(1/2,1/2,\bmu_1, \bmu_2, I_p, I_p): \\
 &\bmu_1=\lambda_1 \bm e_1+\sum_{i=2}^p \frac{\lambda_2}{\sqrt n}\cdot u_i\cdot \bm e_i, \bm u\in \mathcal A_{p,p/4},\bmu_2=\bm 0_p\} ,
 \end{align*} 
 where $ \mathcal A_{p,p/4}$ is defined in Lemma \ref{packing}, and $\lambda_1,\lambda_2$ are of constant order and chosen later. %such that $\Theta_1\subset\Theta_p^{(1)}$. 

According to Lemma \ref{packing}, there is a subset of $\Theta_1$ with logarithm cardinality being of order $p$, such that for any $\bm\theta_{\bm u}, \bm\theta_{\bm u'}$ in this subset, we have $\rho_H(\bm u,\bm u')\ge p/8$.  We are going to apply Lemma \ref{fano} to this subset to complete the proof of Theorem  \ref{general-lb1}.

For $\bm u\in \mathcal A_{p,p/4}$, let $\bmu_{\bm u}=\lambda_1 \bm e_1+\sum_{i=2}^p \frac{\lambda_2}{\sqrt n}\cdot u_i\cdot \bm e_i$. Note that for two multivariate normal distributions $\Pro_{\bth_{\bm u}}=N_p(\bmu_{\bm u},I_p)$ and $\Pro_{\bth_{\bm u'}}=N_p(\bmu_{\bm u'},I_p)$, the KL divergence between them are upper bounded by
 \begin{align*}
 KL(\Pro_{\bth_{\bm u}},\Pro_{\bth_{\bm u'}})=\frac{1}{2}\|\bmu_{\bm u}-\bmu_{\bm u'}\|_2^2\le{\frac{\lambda_2^2\cdot p}{4n}}.
 \end{align*}
To use Lemma \ref{fano} to prove Theorem  \ref{general-lb1}, we further need to show that for any $\bth_{\bm u}, \bth_{\bm u'}$, $$
 [R_{\bth}(G)-R_{\bth}(G^*_{\bth_{\bm u}})]+ [R_{\bth}(G)-R_{\bth}(G^*_{\bth_{\bm u'}})]\gtrsim \frac{p}{n}.
 $$
% $\bm\theta_{\bm u}, \bm\theta_{\bm u'}$, $$ [\Pro_{\bth_{\bm u}}(G(\bm z)\neq L(\bm z))-\Pro_{\bth_{\bm u}}(G^*_{\bth_{\bm u}}(\bm z) \neq L(\bm z))]+ [\Pro_{\bth_{\bm u'}}(G(\bm z)\neq L(\bm z))-\Pro_{\bth_{\bm u'}}(G^*_{\bth_{\bm u'}}(\bm z) \neq L(\bm z))]\gtrsim \frac{p}{n}.$$

By Lemma  \ref{transition} and \ref{triangle}, \begin{align*}
 & [R_{\bth}(G)-R_{\bth}(G^*_{\bth_{\bm u}})]+ [R_{\bth}(G)-R_{\bth}(G^*_{\bth_{\bm u'}})]\\%[\Pro_{\bth_{\bm u}}(G(\bm z)\neq L(\bm z))-\Pro_{\bth_{\bm u}}(G^*_{\bth_{\bm u}}(\bm z) \neq L(\bm z))]+ [\Pro_{\bth_{\bm u'}}(G(\bm z)\neq L(\bm z))-\Pro_{\bth_{\bm u'}}(G^*_{\bth_{\bm u'}}(\bm z) \neq L(\bm z))] \\
 \gtrsim& L^2_{\bth_{\bm u}}( G)+L^2_{\bth_{\bm u'}}( G)\ge \frac{1}{2}(L_{\bth_{\bm u}}( G)+L_{\bth_{\bm u'}}( G))^2\ge \frac{1}{2} (L_{\bth_{\bm u}}(G^*_{\bth_{\bm u'}})-\sqrt{\frac{KL(\Pro_{\bth_{\bm u}},\Pro_{\bth_{\bm u'}})}{2}})^2.
 \end{align*}
 
 Since now that $KL(\Pro_{\bth_{\bm u}},\Pro_{\bth_{\bm u'}})\le\frac{\lambda_2^2\cdot p}{4n}$, it's then sufficient to show  $L_{\bth_{\bm u}}(G^*_{\bth_{\bm u'}})\ge c \sqrt{\frac{p}{n}}$ for some $c>\frac{\lambda_2}{2\sqrt 2}$. 

 Without loss of generality, we assume that the coordinates of $\bm u$ and $\bm u'$ are ordered such that $ u_i=u_i'=1$ for $i=2,...,m_1$, $u_i=1-u_i'=1$ for $i=m_1+1, ..., m_2$, $u_i=1-u_i'=0$ for $i=m_2+1, ..., m_3$ and $u_i=u_i'=0$ for $i=m_3+1,...,p$. We then have $\rho_H(\bm u,\bm u')=m_3-m_1\ge\frac{p}{8}$. 

Recall that when $\Sigma_1=\Sigma_2=I_p$ and $\bmu_2=\bm 0_p$, the oracle rule is given by$$
G^*_{\bth}(\bm z)=1+{\1}\{-\bmu_1^\top(\bm z-\frac{\bmu_1}{2}) >0\}.
$$ 

Then 
%$$
%G^*_{\bth}(\bm x; \bth)=\1\{\frac{1}{\sqrt s} 1_{p,s}^\top \bm x>0\},
%$$
\begin{align*}
&G^*_{\bth_{\bm u}}(\bm z )=1+\1\{-\frac{\lambda_2}{\sqrt n}\left(\sum_{i=2}^{m_1} z_i+\sum_{i=m_1+1}^{m_2} z_i\right)-\lambda_1 z_1+\frac{1}{2}\lambda_1^2+\frac{\lambda_2^2(p-1)}{8n}>0\},
\end{align*}
and
\begin{align*}
G^*_{\bth_{\bm u'}}(\bm z )=1+\1\{-\frac{\lambda_2}{\sqrt n}\left(\sum_{i=2}^{m_1} z_i+\sum_{i=m_2+1}^{m_3} z_i\right)-\lambda_1 z_1+\frac{1}{2}\lambda_1^2+\frac{\lambda_2^2(p-1)}{8n}>0\}.
\end{align*}

Let $Z_1=-\lambda_1 z_1-\frac{\lambda_2}{\sqrt n}\sum_{i=2}^{m_1} z_i+\frac{1}{2}\lambda_1^2+\frac{\lambda_2^2(p-1)}{8n}$, $Z_2=\frac{\lambda_2}{\sqrt n}\sum_{i=m_1+1}^{m_2} z_i$ and $Z_3=\frac{\lambda_2}{\sqrt n}\sum_{i=m_2+1}^{m_3} z_i$, then
$$
G^*_{\bth_{\bm u}}(\bm z )=1+\1\{Z_1-Z_2>0\} \text{ and } G^*_{\bth_{\bm u'}}(\bm z )=1+\1\{Z_1-Z_3>0\}, 
$$and therefore
\begin{align*}
L_{\bth_{\bm u}}(G^*_{\bth_{\bm u'}})=&\Pro_{\bth_{\bm u}}(G^*_{\bth_{\bm u'}}(\bm z)\neq G^*_{\bth_{\bm u}}(\bm z))\\
=&\Pro_{\bth_{\bm u}}(Z_2\le Z_1\le Z_3)+\Pro_{\bth_{\bm u}}(Z_3\le Z_1\le Z_2)\\
\ge&\Pro_{\bth_{\bm u}}(Z_2\le Z_1\le Z_3)\\
=&\frac{1}{2}\Pro_{\bm z\sim N_p(  \bmu_{\bm u},I_p)}\left( Z_2\le Z_1\le Z_3\right)+\frac{1}{2}\Pro_{\bm z\sim N_p(  \bm 0_{p},I_p)}\left( Z_2\le Z_1\le Z_3\right)\\
\ge&\frac{1}{2}\Pro_{\bm z\sim N_p(  \bm 0_{p},I_p)}\left( Z_2\le Z_1\le Z_3\right),%\gtrsim\sqrt\frac{p}{n},
\end{align*}

Then, since $Z_1\sim N\left(\frac{1}{2}\lambda_1^2+\frac{\lambda_2^2(p-1)}{8n},\lambda_1^2+\lambda_2^2p/(4n)\right)$, the density of $Z_1$, $f(z)$ satisfies, $$f(z)\ge\frac{1}{\sqrt{2\pi(\lambda_1^2+\lambda_2^2p/(4n))}}\exp(-\frac{(z-\lambda_1^2/2-\lambda_2^2(p-1)/(8n))^2}{2(\lambda_1^2+\lambda_2^2p/(4n))^2}),$$ leading to  $$%$f(z)\ge\frac{1}{\sqrt{2\pi(\lambda^2+p/(2n))}}\exp(-(z-\lambda^2/2-p/(8n))^2)$, satisfies that $$
f(z)\ge c_1(\lambda_1,\lambda_2), \text{ for } z\in[-\lambda_2\sqrt{p/n},\lambda_2\sqrt{p/n}].
$$
for some constant $c_1(\lambda_1,\lambda_2)=\frac{1}{\sqrt{2\pi(\lambda_1^2+\lambda_2^2p/(4n))}}\exp(-\frac{(\lambda_2\sqrt{p/n}+\lambda_1^2/2+\lambda_2^2(p-1)/(8n))^2}{{2(\lambda_1^2+\lambda_2^2p/(4n))^2}})$.

 In addition, since $m_3-m_1\in(\frac{p}{8},\frac{p}{2})$, $Z_3-Z_2$ is normally distributed with mean $0$ and variance of order $\frac{p}{n}$, and therefore we claim that for some constant $c_2$,$$
 \E[(Z_3-Z_2)\cdot\mathbbm 1\{-\lambda_2\sqrt{\frac{p}{n}}<Z_2<Z_3<\lambda_2\sqrt{\frac{p}{n}}\}]\ge c_2\lambda_2\sqrt\frac{p}{n}.
 $$
 In fact,
\begin{align*}
&\E[(Z_3-Z_2)\cdot\mathbbm 1\{-\lambda_2\sqrt{\frac{p}{n}}<Z_2<Z_3<\lambda_2\sqrt{\frac{p}{n}}\}]\\
\ge&\E[(Z_3-Z_2)\cdot\mathbbm 1\{-\lambda_2\sqrt{\frac{p}{n}}<Z_2<-\frac{\lambda_2}{2}\sqrt{\frac{m_2-m_1}{n}}, \;\;\frac{\lambda_2}{2}\sqrt{\frac{m_3-m_2}{n}}<Z_3<\lambda_2\sqrt{\frac{p}{n}}\}]\\
\ge& \lambda_2\sqrt\frac{p}{n}\cdot \Pro(-\lambda_2\sqrt{\frac{p}{n}}<Z_2<-\frac{\lambda_2}{2}\sqrt{\frac{m_2-m_1}{n}})\cdot\Pro(\frac{\lambda_2}{2}\sqrt{\frac{m_3-m_2}{n}}<Z_3<\lambda_2\sqrt{\frac{p}{n}})\\
\ge& \lambda_2\sqrt\frac{p}{8n}\cdot \Pro_{Z\sim N(0,1)}(-\sqrt{\frac{p}{m_2-m_1}}<Z<-\frac{1}{2})\cdot\Pro_{Z\sim N(0,1)}(\frac{1}{2}<Z<\sqrt{\frac{p}{m_3-m_2}})\\
\ge& \lambda_2\sqrt\frac{p}{8n}\cdot \Pro_{Z\sim N(0,1)}(-\sqrt{2}<Z<-\frac{1}{2})\cdot\Pro_{Z\sim N(0,1)}(\frac{1}{2}<Z<\sqrt{2}):= c_2\lambda_2\sqrt\frac{p}{n},
\end{align*}
where $c_2=\sqrt{\frac{1}{8}}\Pro_{Z\sim N(0,1)}(-\sqrt{2}<Z<-\frac{1}{2})\cdot\Pro_{Z\sim N(0,1)}(\frac{1}{2}<Z<\sqrt{2})$ is of constant order and the inequality above uses $\sqrt{m_2-m_1}+\sqrt{m_3-m_2}\ge \sqrt{m_3-m_1}\ge\sqrt{p/8}$, $m_2-m_1, m_3-m_2\le m_3-m_1\le p/2$.

 Then we have
  \begin{align*}
&\Pro_{\bm z\sim N_p(  \bm 0_{p},I_p)}\left( Z_2\le Z_1\le Z_3\right)\ge\Pro_{\bm z\sim N_p(  \bm 0_{p},I_p)}\left( Z_2\le Z_1\le Z_3, -\lambda_2\sqrt{\frac{p}{n}}<Z_2<Z_3<\lambda_2\sqrt{\frac{p}{n}}\right)\\
=& \E_{Z_2}[\int_{Z_2}^{Z_3} f(z_1) \:dz_1\cdot\1\{-\lambda_2\sqrt{\frac{p}{n}}<Z_2<Z_3<\lambda_2\sqrt{\frac{p}{n}}\}]\\
\ge& c_1(\lambda_1,\lambda_2)\cdot \cdot\E_{Z_2}[(Z_3-Z_2)\cdot\mathbbm 1\{-\lambda_2\sqrt{\frac{p}{n}}<Z_2<Z_3<\lambda_2\sqrt{\frac{p}{n}}\}]\\
\ge & c_1(\lambda_1,\lambda_2)c_2\lambda_2\cdot\sqrt\frac{p}{n}.
\end{align*}

Since $p\lesssim n$, we have $c_1(\lambda_1,\lambda_2)\to\infty$ when $\lambda_1,\lambda_2\to 0$. Therefore, we can choose $\lambda_1,\lambda_2$ to be sufficiently small such that $c_1(\lambda_1,\lambda_2)c_2\lambda_2\sqrt\frac{p}{n}\ge  \frac{\lambda_2}{2\sqrt 2}\sqrt\frac{p}{n}$. 
This completes the proof. 

%%%%%%%%%%%%%%%%%%%%
\subsubsection{Proof of Theorem \ref{general-lb2}}
%%%%%%%%%%%%%%%%%%%%

At first we construct the following least favorable subset, which characterizes the difficulty of the general QDA problem. For simplicity of notation, we use the letters $\lambda_1, \lambda_2$ in this section, whose values are different from those in Section~\ref{sec:pfthm1}.

Since the $KL$-divergence and $\ell_2$ norm are invariant to translations and orthogonal transformations, without loss of generality, we assume that $ \bmu_1^*=-\bmu_2^*=\lambda_1\bm e_1+\tilde\lambda_1\bm e_2$ for some constants $\lambda_1,\tilde\lambda_1>0$ whose values are determined later, with $2\sqrt{\lambda_1^2+\tilde\lambda_1^2}=\|\bmu_1^*-\bmu_2^*\|_2$. In addition, we assume that  $p/4$ is an integer.

Now let's consider
% \begin{align*}
% \Theta_2=\{\bth_{\bm u}=(1/2,1/2,\bmu_1, \bmu_2, \Sigma_1^{\bm u}, \Sigma_2): &\Sigma_1^{\bm u}=\sigma^2  (I_p+\frac{\lambda_2}{2\sqrt n} \sum_{i=2}^{p/4}(2u_i-1)E_{i,i}+\sqrt \frac{2}{{3 p}}\sum_{i={p/4+1}}^{p}E_{i,i})^{-1},\\
%  &\bm u\in \mathcal A_{p/4,s}, \Sigma_2=\sigma^2 I_p,  \bmu_1=-\bmu_2=\lambda\bm e_1 \} ,
% \end{align*} 
 \begin{align*}
 \Theta_2=\{&\bth_{\bm u}=(1/2,1/2,\lambda_1\bm e_1+\tilde\lambda_1\bm e_2, -\lambda_1\bm e_1-\tilde\lambda_1\bm e_2, \Sigma_1^{\bm u}, \Sigma_2): \\
 &\Sigma_1^{\bm u}= (I_p+\tilde\lambda_2 E_{2,2}+\frac{\lambda_2}{\sqrt n} \sum_{i=3}^{p/2}u_iE_{i,i})^{-1},\bm u\in \mathcal A_{p,p/4}, \Sigma_2= I_p+\tilde\lambda_2 E_{2,2}   \} ,
 \end{align*} 
 where $ \mathcal A_{p,p/4}$ is defined in Lemma \ref{packing} . 
 
 According to Lemma \ref{packing}, there is a subset of $\Theta_1$ with logarithm cardinality being of order $p$, such that for any $\bm\theta_{\bm u}, \bm\theta_{\bm u'}$ in this subset, we have $\rho_H(\bm u,\bm u')\ge p/8$.  We are going to apply Lemma \ref{fano} to this subset to complete the proof of Theorem  \ref{general-lb2}.

 At first we note that for two multivariate normal distribution $N_p(\bm\mu_1^*,\Sigma_1^{\bm u})$ and $N_p(\bm\mu_1^*,\Sigma_1^{\bm u'})$, using the fact that $\log(1+x)\asymp x-x^2/2+o(x^2)$ for $x=o(1)$, the KL divergence between them are upper bounded by  \begin{align*}
 KL&=\frac{1}{2}\left[\log\frac{|\Sigma_1^{\bm u'}|}{|\Sigma_1^{\bm u}|}-p+\tr((\Sigma_1^{\bm u'})^{-1}\Sigma_1^{\bm u})\right]\\
 &=\frac{1}{2}\left[\sum_{i=3}^{p}\log\frac{1+\frac{\lambda_2}{\sqrt n}u_i'}{1+\frac{\lambda_2}{\sqrt n}u_i}-\rho_H(\bm u, \bm u')+\sum_{i=3}^{p}\frac{1+\frac{\lambda_2}{\sqrt n}u_i}{1+\frac{\lambda_2}{\sqrt n}u_i'}\right]\\
  &=\frac{1}{2}\left[-\sum_{i=3}^{p}\log\left(1+\frac{\frac{\lambda_2}{\sqrt n}(u_i-u_i')}{1+\frac{\lambda_2}{\sqrt n}u_i'}\right)+\sum_{i=3}^{p}\frac{\frac{\lambda_2}{\sqrt n}(u_i-u_i')}{1+\frac{\lambda_2}{\sqrt n}u_i'}\right]\\
  &= \frac{1}{4}\sum_{i=3}^{p} \frac{1}{n} (u_i-u_i')^2+o(\frac{p}{n})\le\frac{\lambda_2^2p}{16n}+o(\frac{p}{n})\le\frac{\lambda_2^2p}{8n}.
 \end{align*}

Therefore we have $KL(\Pro_{\bth_{\bm u}}, \Pro_{\bth_{\bm u'}})\le  \lambda_2^2p/{(8n)}$. To use Lemma \ref{fano} to prove Theorem  \ref{general-lb2}, we further need to show that for any $\bm\theta_{\bm u}, \bm\theta_{\bm u'}$, 
$$
 [R_{\bth}(G)-R_{\bth}(G^*_{\bth_{\bm u}})]+ [R_{\bth}(G)-R_{\bth}(G^*_{\bth_{\bm u'}})]\gtrsim \frac{p}{n}.
$$
By Lemma \ref{transition} and  \ref{triangle}, \begin{align*}
 & [R_{\bth}(G)-R_{\bth}(G^*_{\bth_{\bm u}})]+ [R_{\bth}(G)-R_{\bth}(G^*_{\bth_{\bm u'}})] \\
 \ge& L^2_{\bth_{\bm u}}(\hat G)+L^2_{\bth_{\bm u'}}(\hat G)\ge \frac{1}{2}(L_{\bth_{\bm u}}(\hat G)+L_{\bth_{\bm u'}}(\hat G))^2\ge \frac{1}{2} (L_{\bth_{\bm u}}(G^*_{\bth_{\bm u'}})-\sqrt{\frac{KL(\Pro_{\bth_{\bm u}},\Pro_{\bth_{\bm u'}})}{2}})^2.
 \end{align*}
 
 Since now that $KL(\Pro_{\bth_{\bm u}},\Pro_{\bth_{\bm u'}})\le\lambda_2^2\frac{p}{8n}$, it's then sufficient to show  $L_{\bth_{\bm u}}(G^*_{\bth_{\bm u'}})\ge c \sqrt{\frac{p}{n}}$ for some $c>\lambda_2/4$.

Recall that
\begin{equation*}
G^*_{\bth}(\bm z)={\1}\{(\bm z-\bmu_1)^\top D(\bm z-\bmu_1)-2\bdel^\top\Omega_2(\bm z-\bmu_1)+\bdel^\top\Omega_2\bdel-\log({|\Sigma_1|\over |\Sigma_2|}) >0\},
\end{equation*} where $ \bdel=\bmu_2-\bmu_1$, $D=\Omega_2-\Omega_1$. 

 Without loss of generality, we assume that $ u_i=u_i'=1$ when $i=3,...,m_1$, $u_i=1-u_i'=1$ when $i=m_1+1, ..., m_2$, $u_i=1-u_i'=0$ when $i=m_2+1, ..., m_3$ and $u_i=u_i'=0$ when $i=m_3+1,...,p$.

Then with a little abuse of notation, we have $\bm z\sim \frac{1}{2}N_p(\bmu_1,  \Sigma_1^{\bm u})+\frac{1}{2}N_p(\bmu_2, \Sigma_2)$ with $\bmu_!1-\bmu_2=\lambda_1\bm e_1+\tilde\lambda_1\bm e_2$. Using the fact that $\log(1+\frac{\lambda_2}{\sqrt n})= \frac{\lambda_2}{\sqrt n}-\frac{\lambda_2^2}{2n}+o(\frac{1}{n})$, we have
%$$
%G^*_{\bth}(\bm x; \bth)=\1\{\frac{1}{\sqrt s} 1_{p,s}^\top \bm x>0\},
%$$
 \fontsize{10pt}{9}\selectfont

\begin{align*}
&G^*_{\bth_{\bm u}}(\bm z )=1+\1\{\frac{\lambda_2}{\sqrt n}\left(\sum_{i=3}^{m_1} (z_i^2-1)+\sum_{i=m_1+1}^{m_2} (z_i^2-1)\right)+4\lambda_1 z_1+4\frac{\tilde\lambda_1}{1+\tilde\lambda_2}z_2+\frac{p}{8n}+o(\frac{p}{n})>0\},\\%+\sqrt \frac{2}{{3 p}}\sum_{i=p/4+1}^{p} (z_i^2-1)\\
%&\quad+4\lambda z_1+O(\frac{1}{n})>0\},
\end{align*}
and
\begin{align*}
&G^*_{\bth_{\bm u'}}(\bm z )=1+\1\{\frac{\lambda_2}{\sqrt n}\left(\sum_{i=3}^{m_1} (z_i^2-1)+\sum_{i=m_2+1}^{m_3} (z_i^2-1)\right)+4\lambda_1 z_1+4\frac{\tilde\lambda_1}{1+\tilde\lambda_2}z_2+\frac{p}{8n}+o(\frac{p}{n})>0\}.%+\sqrt \frac{2}{{3 p}}\sum_{i=p/4+1}^{p} (z_i^2-1)\\
%&\quad+4\lambda z_1+O(\frac{1}{n})>0\}.
\end{align*}

\normalsize
Let $Z_1=-(4\lambda_1 z_1+4\frac{\tilde\lambda_1}{1+\tilde\lambda_2}z_2+\frac{\lambda_2}{\sqrt n}\sum_{i=3}^{m_1} (z_i^2-1)+\frac{p}{8n})$%+\sqrt \frac{2}{{3 p}}\sum_{i=p/4+1}^{p} (z_i^2-1)$
, $Z_2=\frac{\lambda_2}{\sqrt n}\sum_{i=m_1+1}^{m_2} (z_i^2-1)$, $Z_3=\frac{\lambda_2}{\sqrt n}\sum_{i=m_2+1}^{m_3} (z_i^2-1)$, then
$$
G^*_{\bth_{\bm u}}(\bm z )=\1\{-Z_1+Z_2+o(\frac{p}{n})>0\} \text{ and } G^*_{\bth_{\bm u'}}(\bm z )=\1\{-Z_1+Z_3+o(\frac{p}{n})>0\}, 
$$
and
\begin{align*}
L_{\bth_{\bm u}}(G^*_{\bth_{\bm u'}})=&\Pro_{\bth_{\bm u}}(G^*_{\bth_{\bm u'}}(\bm z)\neq G^*_{\bth_{\bm u}}(\bm z))\\
\ge&\frac{1}{2}\Pro_{\bm z\sim N_p(  \bm \mu_{1},\Sigma_1^{\bm u})}\left( Z_2+o(\frac{p}{n})\le Z_1\le Z_3+o(\frac{p}{n})\right)\\
&+\frac{1}{2}\Pro_{\bm z\sim N_p(  \bm \mu_{2},\Sigma_2)}\left( Z_3+o(\frac{p}{n})\le Z_1\le Z_2+o(\frac{p}{n})\right)\\
\ge&\frac{1}{2}\Pro_{\bm z\sim N_p(  \bm \mu_{1},\Sigma_2)}\left( Z_2\le Z_1\le Z_3\right)+o(\frac{p}{n}).
\end{align*}

By central limit theorem, $\frac{\sqrt{n}}{\lambda_2\sqrt{m_2-m_1}}Z_2$, $\frac{\sqrt{n}}{\lambda_2\sqrt{m_3-m_2}}Z_3$ converges to the standard normal distribution $N(0,1)$. Since $m_3-m_2=\rho_H(\bm u, \bm u')\ge p/8$, and $\lim\sup_{n,p\to\infty} \frac{p}{n}\le C_1$, similar as the derivation in Section~\ref{sec:pfthm1}, there exists a constant $c_2$, such that $n,p$ are sufficiently large, 
\begin{align*}
&\E[(Z_3-Z_2)\cdot\mathbbm 1\{-\lambda_2\sqrt{\frac{p}{n}}<Z_2<Z_3<\lambda_2\sqrt{\frac{p}{n}}\}]\\
\ge&\E[(Z_3-Z_2)\cdot\mathbbm 1\{-\lambda_2\sqrt{\frac{p}{n}}<Z_2<-\frac{\lambda_2}{2}\sqrt{\frac{p}{n}}, \;\;\frac{\lambda_2}{2}\sqrt{\frac{p}{n}}<Z_3<\lambda_2\sqrt{\frac{p}{n}}\}]\\
\ge& \lambda_2\sqrt\frac{p}{n}\cdot \Pro(-\lambda_2\sqrt{\frac{p}{n}}<Z_2<-\frac{\lambda_2}{2}\sqrt{\frac{m_2-m_1}{n}})\cdot\Pro(\frac{\lambda_2}{2}\sqrt{\frac{m_3-m_1}{n}}<Z_3<\lambda_2\sqrt{\frac{p}{n}})\\
\ge& \lambda_2\sqrt\frac{p}{8n}\cdot \Pro_{Z\sim N(0,1)}(-\sqrt{\frac{p}{m_2-m_1}}<Z<-\frac{1}{2})\cdot\Pro_{Z\sim N(0,1)}(\frac{1}{2}<Z<\sqrt{\frac{p}{m_3-m_2}})\\
\ge& \lambda_2\sqrt\frac{p}{8n}\cdot \Pro_{Z\sim N(0,1)}(-\sqrt{2}<Z<-\frac{1}{2})\cdot\Pro_{Z\sim N(0,1)}(\frac{1}{2}<Z<\sqrt{2})\ge c_2\lambda_2\sqrt\frac{p}{n}.
\end{align*}
%$$
%\E_{Z_2}[(Z_2-Z_3)\cdot\mathbbm 1\{-\lambda_2\sqrt{\frac{p}{n}}<-Z_2<-Z_3<\lambda_2\sqrt{\frac{p}{n}}\}]\ge c_1\lambda_2\sqrt\frac{p}{n}.
%$$

Similar to that in Section~\ref{sec:pfthm1}, let's denote the probability density function of $Z_1$ by $f$. Use central limit theorem again, when $\bm z\sim N_p(  \bm \mu_{1},\Sigma_2)$, $p\lesssim n$, and $n,p$ are sufficiently large, $Z_1\approx N(-4\lambda_1^2-\frac{4\tilde\lambda_1^2}{1+\tilde\lambda_2}+\frac{p}{8n}, \lambda_1^2+\frac{\tilde\lambda_1^2}{1+\tilde\lambda_2}+\frac{2(m_1-2)\lambda_2^2}{n})$ if $m_1\to\infty$. Therefore, there exists constant $c_1(\lambda_1,\tilde\lambda_1, \lambda_2,\tilde\lambda_2)$, such that $\inf_{|x|<\lambda_2\sqrt{p/n}} f(x)>c_1(\lambda_1,\tilde\lambda_1, \lambda_2,\tilde\lambda_2)$, and $c_1(\lambda_1,\tilde\lambda_1, \lambda_2,\tilde\lambda_2)$ goes to infinity when $\lambda_1,\lambda_2\to0,  \tilde\lambda_2\to\infty$, and $\tilde\lambda_1$ is chosen such that $\sqrt{\lambda_1^2+\tilde\lambda_1^2}=\|\bmu_1^*-\bmu_2^*\|_2/2$.
 \begin{align*}
&\Pro_{\bm z\sim N_p(  \bm \mu_{1},\Sigma_2)}\left( Z_2\le Z_1\le Z_3\right)\ge\Pro_{\bm z\sim N_p(  \bm \mu_{1},\Sigma_2)}\left( Z_2\le Z_1\le Z_3, -\lambda_2\sqrt{\frac{p}{n}}<Z_2<Z_3<\lambda_2\sqrt{\frac{p}{n}}\right)\\
=& \E_{Z_2}[\int_{Z_2}^{Z_3} f(z_1) \:dz_1\cdot\1\{-\lambda_2\sqrt{\frac{p}{n}}<Z_2<Z_3<\lambda_2\sqrt{\frac{p}{n}}\}]\\
\ge& c_1(\lambda_1,\tilde\lambda_1, \lambda_2,\tilde\lambda_2)\cdot \cdot\E_{Z_2}[(Z_3-Z_2)\cdot\mathbbm 1\{-\lambda_2\sqrt{\frac{p}{n}}<Z_2<Z_3<\lambda_2\sqrt{\frac{p}{n}}\}]\\
\ge & c_1(\lambda_1,\tilde\lambda_1, \lambda_2,\tilde\lambda_2)c_2\lambda_2\cdot\sqrt\frac{p}{n}.
\end{align*}

%Since $\sqrt{\lambda_1^2+\tilde\lambda_1^2}=\|\bmu_1^*-\bmu_2^*\|_2/2$, and $2\sqrt{\lambda_1^2+\tilde\lambda_1^2}=\|\bmu_1^*-\bmu_2^*\|_2$ goes to infinity when $\lambda_1,\lambda_2\to0$ and $\tilde\lambda_2\to\infty$,

Therefore, by choosing sufficiently small $\lambda_1, \lambda_2$ and large $\tilde\lambda_2$ (doesn't depend on $n,p$), we have $c_2 c_1(\lambda_1,\tilde\lambda_1, \lambda_2,\tilde\lambda_2)\cdot\lambda_2\sqrt\frac{p}{n}\ge\frac{\lambda_2}{4}\sqrt\frac{p}{n}$.
\qed

%%%%%%%%%%%%%%%%%%%%
\subsection{Proof of the Theorem \ref{estimation}}
%%%%%%%%%%%%%%%%%%%%
%Without loss of generality, we assume $n_1=n_2$, $\pi_1=\pi_2=1/2$ in the proof. 
 To prove Theorem~\ref{estimation} we begin by collecting a few important technical lemmas that will be used in the main proofs.% of the main results.
\subsubsection{Auxiliary Lemmas}

\begin{lemma}\label{gammas:con}
Suppose $\bm X_1, ..., \bm X_n$ $i.i.d.$ $\sim N_p(\bmu, \Sigma)$, and assume that $\hat \bmu$, $\hat\Sigma$ are the sample mean and sample covariance matrix respectively. Let $\Gamma(s;p)=\{\bm u\in\R^p: \|\bm u\|_2=1, \|\bm u_{S^C}\|_1\le \|\bm u_{S}\|_1, \text{ for some }S\subset[p] \text{ with }|S|=s\}$, then with probability at least $1-p^{-1}$, 
$$
\sup_{\bm u\in \Gamma(s; p)}\bm u^\top(\hat\bmu-\bmu)\lesssim \sqrt{\frac{s\log p}{n}};%\quad \sup_{i\in [p]}\bm e_i^\top(\hat\Sigma-\Sigma)\bm b\lesssim \sqrt{\frac{\log p}{n}};
$$
$$
\sup_{\bm u,\bm v\in \Gamma(s; p)}\bm u^\top(\hat\Sigma-\Sigma)\bm v\lesssim \sqrt{\frac{s\log p}{n}};\quad \sup_{\bm a\in \Gamma(s; p^2)}\bm a^\top\vect(\hat\Sigma-\Sigma)\lesssim \sqrt{\frac{s\log p}{n}}.
$$
\end{lemma}

\begin{lemma}\label{gammas:con0}
Suppose $\bm X_1, ..., \bm X_{n_1}$ $i.i.d.$ $\sim N_p(\bmu_1, \Sigma_1)$, $\bm Y_1, ..., \bm Y_{n_2}$ $i.i.d.$ $\sim N_p(\bmu_2, \Sigma_2)$, $n=\min(n_1,n_2)$ and assume that $\hat \bmu_1, \hat\bmu_2$, $\hat\Sigma_1,\hat\Sigma_2$ are the sample means and sample covariance matrices. Denote $V=\frac{1}{2}\Sigma_1 \otimes \Sigma_2+\frac{1}{2}\Sigma_2 \otimes \Sigma_1$ and $\hat V=\frac{1}{2}\hat\Sigma_1 \otimes \hat\Sigma_2+\frac{1}{2}\hat\Sigma_2 \otimes \hat\Sigma_1$. Assume that $\bbeta=\Omega_2(\bmu_2-\bmu_1)$ and $\vect(D) $has bounded $\ell_2$ norm, then with probability at least $1-p^{-1}$, 
$$
\|\hat\bmu_k-\bmu_k\|_\infty\lesssim \sqrt{\frac{\log p}{n}},
\quad\|(\hat\Sigma_k-\Sigma_k)\bbeta\|_\infty\lesssim \sqrt{\frac{\log p}{n}}, \quad k=1,2;
$$
$$
\|\vect(\hat\Sigma-\Sigma)\|_\infty\lesssim \sqrt{\frac{\log p}{n}};\quad \|(\hat V-V)\vect(D)\|_\infty\lesssim \sqrt{\frac{\log p}{n}}.
$$
\end{lemma}

\begin{lemma}
\label{l1-cone}
Suppose $\bm x, \bm y\in\R^p$. Let $\bm h=\bm x-\bm y$. Denote $\mathcal S=\supp(\bm y)$ and $s=|S|$. If $\|\bm x\|_1\le\|\bm y\|_1$, then $h\in \Gamma(s;p)$, that is, $$
\|\bm h_{\mathcal S^c}\|_1\le\|\bm h_S\|_1. 
$$
\end{lemma}

\begin{lemma}\label{lemma:logdet}
For any two matrices $A,B\in\R^{p\times p}$ with non-negative eigenvalues, $$
\big|\log|A|-\log|B|\big|\le \max\{| \tr(B^{-1}(A-B))|,  |\tr(A^{-1}(B-A))| \}.
$$
\end{lemma}
%At last, to prove the upper results Theorem \ref{estimation} and \ref{Rn}, the following lemmas are needed. Lemma \ref{lemma:logdet} is used for deriving the consistency of estimating the log determinant term in $Q(\bm z;\bth)$ in \eqref{Qz}.  Lemma \ref{l1-cone} is used for transforming the $\ell_\infty$ consistency in Theorem \ref{estimation} to consistencies under $\ell_2$ norm and Frobenius norm. 

\subsubsection{Main proofs}
We prove the consistency of estimation of $D$ first. The consistency of estimating $\bbeta$ can be derived similarly. 

Recall that
\begin{equation}
\hat D=\arg\min_{D\in\R^{p\times p}} \left\{|D|_1:\, \|(\frac{1}{2}\hat\Sigma_1 \otimes \hat\Sigma_2+\frac{1}{2}\hat\Sigma_2 \otimes \hat\Sigma_1)\vect(D)-\vect(\hat\Sigma_1)+\vect(\hat\Sigma_2)\|_{\infty}\le\lambda_{1,n}\right\}
\label{opt:proof}.
\end{equation}

By Lemma \ref{gammas:con0}, $D$ is a feasible solution to \eqref{opt:proof} with $\lambda_{1,n}=c_1\sqrt{\frac{\log p}{n}}$ when $c_1$ is a sufficiently large constant. Then using Lemma~\ref{l1-cone}, we have $\vect(D-\hat D)\in \Gamma(s_1;p^2)$.

Denote $ V=\frac{1}{2}\Sigma_1 \otimes \Sigma_2+\frac{1}{2}\Sigma_2 \otimes \Sigma_1$, $\bm v_{\Sigma}=\vect(\Sigma_1)-\vect(\Sigma_2)$ and $\hat V=\frac{1}{2}\hat\Sigma_1 \otimes \hat\Sigma_2+\frac{1}{2}\hat\Sigma_2 \otimes \hat\Sigma_1$, $\widehat{\bm v_{\Sigma}}=\vect(\hat\Sigma_1)-\vect(\hat\Sigma_2)$.

We have 
\begin{align*}
V\vect(D)=&(\frac{1}{2}\Sigma_1 \otimes \Sigma_2+\frac{1}{2}\Sigma_2 \otimes \Sigma_1)\vect(D)=\vect(\frac{1}{2}\Sigma_1 D \Sigma_2+\frac{1}{2}\Sigma_2 D \Sigma_1)\\
=&\vect(\Sigma_1-\Sigma_2)=\bm v_{\Sigma}.
\end{align*}

In addition, over the parameter space $\Theta_p(s_1, s_2)$, $$
\|V^{-1}\|_2=\|\Omega_1\otimes\Omega_2\|_2=\|\Omega_1\|_2\cdot\|\Omega_2\|_2\le M_1^2. $$
which is followed by $\lambda_{\min}(V)\ge M_1^{-2}.$ 

As a consequence, by Lemma~\ref{gammas:con}, with probability at least $1-3p^{-1}$,
\begin{equation}
\begin{aligned}
&|(\vect(\hat D)-\vect(D))^\top  V (\vect(\hat D)-\vect(D))|\\
\le& |(\vect(\hat D)-\vect(D))^\top(\hat V \vect(\hat D)-\widehat{\bm v_{\Sigma}})|+|(\vect(\hat D)-\vect(D))^\top(\hat V -V) \vect(\hat D))|\\
&+|(\vect(\hat D)-\vect(D))^\top(\bm v_{\Sigma}-\widehat{\bm v_{\Sigma}})|\\
\lesssim&\sqrt{s_1}\|\vect(\hat D)-\vect(D)\|_2\cdot \|\hat V \vect(\hat D)-\widehat{\bm v_{\Sigma}}\|_\infty\\
&+\|\vect(\hat D)-\vect(D)\|_2\cdot\sqrt\frac{s_1\log p}{n}\cdot\|\vect(D)-\vect(\hat D)\|_2\\
&+\|\vect(\hat D)-\vect(\hat D)\|_2\sqrt\frac{s_1\log p}{n}\cdot\|\vect(D)\|_2+\|\vect(D)-\vect(\hat D)\|_2 \sqrt\frac{s_1\log p}{n}.
%\lesssim&\sqrt{s}\|\vect(\hat D)-\bbeta\|_2\cdot\sqrt\frac{\log p}{n}%+\|\hat \bbeta_{\rm AdaLDA}-\bbeta\|_2\cdot\sqrt\frac{s\log p}{n}(\|\bbeta\|_2+\|\bbeta-\hat \bbeta_{\rm AdaLDA}\|_2)
\end{aligned}
\end{equation}

In addition, since $|(\vect(\hat D)-\vect(D))^\top  V (\vect(\hat D)-\vect(D))|\ge\lambda_{\min}(V)\|\vect(\hat D)-\vect(D)\|_2^2\ge M_1^{-2}\|\vect(\hat D)-\vect(D)\|_2^2,$
we then have 
\begin{align*} 
\|D-\hat D\|_F=\|\vect(\hat D)-\vect(D)\|_2\lesssim&\sqrt{\frac{s_1\log p}{n}}.
\end{align*}

The estimation error of $\bbeta$ can be derived similarly. 
%Recall that 
%\begin{equation*}
%\hat\bbeta=\arg\min_{\bbeta\in\R^{p}}\|\bbeta\|_1:\,\left\{ \| \hat\Sigma_2 \bbeta -\hat\bmu_2+\hat\bmu_1\|_{\infty}\le\lambda_{2,n}\right\},
%\end{equation*}
%and $\hat\bdel=\hat\bmu_2-\hat\bmu_1$. 
 By Lemma \ref{gammas:con0}, $\bbeta$ is a feasible solution to \eqref{opt2} with $\lambda_{2,n}=c_2\sqrt{\frac{\log p}{n}}$ when $c_2$ is sufficiently large. Then using Lemma~\ref{l1-cone}, we have $\bbeta-\hat\bbeta\in \Gamma(s_2;p)$.

Then with probability at least $1-3p^{-1}$,
\begin{equation}
\begin{aligned}
&|(\hat \bbeta_{ }-\bbeta)^\top\Sigma_2 (\hat \bbeta_{ }-\bbeta)|\\
\le& |(\hat \bbeta_{ }-\bbeta)^\top(\hat\Sigma_2 \hat \bbeta_{ }-\hat\bdel)|+|(\hat \bbeta_{ }-\bbeta)^\top(\hat\Sigma_2 -\Sigma_2) \hat\bbeta)|+|(\hat \bbeta_{ }-\bbeta)^\top(\bdel-\hat\bdel)|\\
\lesssim&\sqrt{s_2}\|\hat\bbeta-\bbeta\|_2\cdot \|\hat\Sigma \hat \bbeta-\hat\bdel\|_\infty+\|\hat\bbeta-\bbeta\|_2\cdot\sqrt\frac{s_2\log p}{n}\cdot\|\bbeta-\hat\bbeta\|_2\\
&+\|\bbeta-\hat\bbeta\|_2\sqrt\frac{s_2\log p}{n}\cdot\|\bbeta\|_2+\|\bbeta-\hat\bbeta\|_2 \sqrt\frac{s_2\log p}{n}.
%\lesssim&\sqrt{s}\|\hat \bbeta_{ }-\bbeta\|_2\cdot\sqrt\frac{\log p}{n}%+\|\hat \bbeta_{\rm AdaLDA}-\bbeta\|_2\cdot\sqrt\frac{s\log p}{n}(\|\bbeta\|_2+\|\bbeta-\hat \bbeta_{\rm AdaLDA}\|_2)
\end{aligned}
\end{equation}

Similarly, since $\lambda_{\min}(\Sigma_2)\ge M_1^{-1}$, we  have with probability at least $1-p^{-1}$, $$
\|\bbeta-\hat\bbeta\|_2\lesssim\sqrt\frac{s_2\log p}{n}.
$$

%%%%%%%%%%%%%%%%%%%%
\subsection{Proof of Theorem \ref{Rn}}\label{sec:pf:ub}
%%%%%%%%%%%%%%%%%%%%

%\begin{theorem}
%Suppose {\bf (C1)-(C4)}  hold, and $ s_1+s_2\lesssim \frac{{n}}{\log p\log^2 n}$. Then the the proposed SDAR classification rule in \eqref{SDAR} satisfies with probability $1-o(1)$,
%$$
%R(\hat G_{\rm SDAR})-R_{\bth}(G^*_{\bth})]\lesssim  (s_1{\frac{\log p}{n}}+s_2{\frac{\log p}{n}})\cdot\log^2 n.
%$$ 
%\end{theorem}
We note here that the notation $c,C$ denote generic constants and their values might vary line by line.
 Recall that the QDA rule is $$1+{\1}\{(\bm z-\bmu_1)^\top D(\bm z-\bmu_1)-2\bbeta^\top(\bm z-\bar \bmu)-\log(|D\Sigma_1+I_p|)+2\log(\frac{\pi_1}{\pi_2}) >0\}.
$$

Let $\bar\bmu=(\bmu_1+\bmu_2)/2$, $Q(\bm z)=(\bm z-\bmu_1)^\top D(\bm z-\bmu_1)-2\bbeta^\top(\bm z-\bar \bmu)-\log(|D\Sigma_1+I_p|)+2\log(\frac{\pi_1}{\pi_2})$, $\hat Q(\bm z)=(\bm z-\hat\bmu_1)^\top \hat D(\bm z-\hat \bmu_1)-2\hat \bbeta^\top(\bm z-\frac{\hat \bmu_1+\hat \bmu_2}{2})-\log(|\hat D\hat \Sigma_1+I_p|)+\log(\frac{\hat\pi_1}{\hat\pi_2})$, and  $M(\bm z)=Q(\bm z)-\hat Q(\bm z)$, we are going to show that there exist some constants $c,C>0$, such that for any $M>0$, $$
\Pro_{\bm z\sim N_p(\bmu_1,\Sigma_1)}\left(|M(\bm z)|>M \sqrt\frac{(s_1+s_2)\log p}{n}\right)\le e^{-cM}+{ Cp^{-1}},
$$
note that the above probability is taken with respect to the random samples $\bm X_1, ..., \bm X_{n_1}$ $i.i.d.$ $\sim N_p(\bmu_1, \Sigma_1)$, $\bm Y_1, ..., \bm Y_{n_2}$ $i.i.d.$ $\sim N_p(\bmu_2, \Sigma_2)$, and $\bm z\sim N_p(\bmu_1,\Sigma_1)$. We will later see how we reduce the mixed distribution of the test sample to the single distribution when we calculate the classification error. 

 Rewrite the QDA rule as $${\1}\{(\bm z-\bmu_1)^\top D(\bm z-\bmu_1)-2\bbeta^\top(\bm z-\bmu_1)+\bbeta^\top({\bmu_2-\bmu_1})-\log(|D\Sigma_1+I_p|)+2\log(\frac{\pi_1}{\pi_2}) >0\}.
$$

We firstly bound the estimation error of the constant term $\bbeta^\top({\bmu_2-\bmu_1})$.
 We have {with probability at least $1-p^{-1}$,}\begin{align*}
&|\bbeta^\top(\bmu_2-\bmu_1)-\hat\bbeta^\top(\hat\bmu_2-\hat\bmu_1)|\le |\hat\bbeta^\top(\bmu_2-\bmu_1-\hat\bmu_2+\hat\bmu_1)|+\|(\hat\bbeta-\bbeta)^\top (\bmu_2-\bmu_1)\|_2\\
\le& \|\hat\bbeta\|_1\cdot \|\bmu_2-\bmu_1-\hat\bmu_2+\hat\bmu_1\|_\infty+\|\hat\bbeta-\bbeta\|_2 \|\bmu_2-\bmu_1\|_2\\
\le&  \|\bbeta\|_1\cdot \|\bmu_2-\bmu_1-\hat\bmu_2+\hat\bmu_1\|_\infty+\|\hat\bbeta-\bbeta\|_2 \|\bmu_2-\bmu_1\|_2\\
\le&  \sqrt s_2\|\bbeta\|_2\cdot \|\bmu_2-\bmu_1-\hat\bmu_2+\hat\bmu_1\|_\infty+\|\hat\bbeta-\bbeta\|_2 \|\bmu_2-\bmu_1\|_2\lesssim \sqrt{\frac{s_2\log p}{n}}.
\end{align*}
For $\log|D\Sigma_1+I_p|$, notice that $D\Sigma_1+I_p=\Omega_2\Sigma_1$ and the product of two positive semidefinite and symmetric matrices has non-negative eigenvalues, followed by $(D\Sigma_1+I_p)^{-1}=\Omega_1\Sigma_2=(\Omega_2-D)\Sigma_2=I_p-D\Sigma_2$, then
\begin{align}\label{ineq:trace}
\notag&\log|D\Sigma_1+I_p|-\log|\hat D\hat\Sigma_1+I_p|\le\tr((D\Sigma_1+I_p)^{-1}(D\Sigma_1-\hat D\hat\Sigma_1))\\
\notag=&\tr((-D\Sigma_2+I_p)(D\Sigma_1-\hat D\hat\Sigma_1))\\
\notag=&\tr((-D\Sigma_2)(D\Sigma_1-\hat D\hat\Sigma_1))+\tr(D\Sigma_1-\hat D\hat \Sigma_1)\\
\notag\le&\|D\Sigma_2\|_F\cdot\|D\Sigma_1-\hat D\hat\Sigma_1\|_F+\tr(D\Sigma_1-\hat D\hat \Sigma_1)\\
\notag\le&\|D\|_F\|\Sigma_2\|_2\cdot\|D\Sigma_1-\hat D\hat\Sigma_1\|_F+\tr(D\Sigma_1-\hat D\hat \Sigma_1)\\
\le&\|D\|_F\|\Sigma_2\|_2\cdot\|D\Sigma_1-\hat D\hat\Sigma_1\|_F+|\tr(\hat D\Sigma_1-\hat D\hat \Sigma_1)|+\tr(D\Sigma_1-\hat D \Sigma_1).
\end{align}

In addition, {with probability at least $1-p^{-1}$,}%define $\|_2\cdot\|_{2,s}$ norm as $\|\Sigma-\hat\Sigma\|_{2,s}=\sup_{\bm u,\bm v\in \Gamma(s; p)}\bm u^\top(\hat\Sigma-\Sigma)\bm v$, then
 \begin{align*}
&\|D\Sigma_1-\hat D\hat\Sigma_1\|_F\le\|D\Sigma_1-\hat D\Sigma_1\|_F+\|\hat D(\Sigma_1-\hat \Sigma_1)\|_F\\
\le&\|\Sigma_1\|_2\|D-\hat D\|_F+\|\Sigma_1-\hat\Sigma_1\|_{2,s_1}\|\hat D\|_F\\
\lesssim&\sqrt{\frac{s_1\log p}{n}}+\|\Sigma_1-\hat\Sigma_1\|_{2,s_1}(\|D\|_F+\sqrt\frac{s_1\log p}{n})\\
\le&\sqrt{\frac{s_1\log p}{n}}+\sqrt{\frac{s_1\log p}{n}}(\|D\|_F+\sqrt\frac{s_1\log p}{n})
%\le&\sqrt{\frac{s_2\log p}{n}}+\sqrt{s_2\cdot\|\hat D\|_1^2\cdot\|\Sigma_1-\hat \Sigma_1\|_\infty\cdot\|\Sigma_1-\hat \Sigma_1\|_\infty}\\
\lesssim\sqrt{\frac{s_1\log p}{n}},
\end{align*}
where $\|\Sigma_1-\hat\Sigma_1\|_{2,s_1}$ is defined as \begin{align*}
\|\Sigma_1-\hat\Sigma_1\|_{2,s_1}&:=\sup_{\|\bm u\|_0\le s_1, \|\bm u\|_2=1}\|(\Sigma_1-\hat\Sigma_1) \bm u\|_2\lesssim\sqrt\frac{s_1\log p}{n},
\end{align*}
where the last inequality is similarly proved as Lemma~\ref{gammas:con}, by using the packing number argument. 

In addition,  {with probability at least $1-p^{-1}$}, $$
|\tr(\hat D\Sigma_1-\hat D\hat \Sigma_1)|\le \sqrt s_1|\Sigma_1-\hat\Sigma_1|_{\infty}\|\hat D\|_F\lesssim\sqrt{\frac{s_1\log p}{n}}.
$$

There is still a remaining term $\tr(D\Sigma_1-\hat D \Sigma_1)$ in \eqref{ineq:trace}, we will leave it there and use it when we derive the distribution of the term involving $\bm z$. The other direction, the upper bound of $\tr(D\Sigma_1-\hat D \Sigma_1)-(\log|D\Sigma_1+I_p|-\log|\hat D\hat\Sigma_1+I_p|)$, can be derived similarly.
%, by using the fact that $\lambda_{min}(A^{-1})\ge c_1$, and $\|A-\hat A\|_2\le\|A-\hat A\|_F\le\sqrt{\frac{s_2\log p}{n}}$, and $$
%\|(\hat A)^{-1}-A^{-1}\|_F=\|(\hat A)^{-1}(\hat A-A)A^{-1}\|_F\le \|(\hat A)^{-1}\|_2\|\hat A-A\|_F \|A^{-1}\|_2.
%$$
Therefore by symmetry, we have {with probability at least $1-p^{-1}$}$$
\left|(\log|D\Sigma_1+I_p|-\log|\hat D\hat\Sigma_1+I_p|)-(\tr(D\Sigma_1-\hat D \Sigma_1))\right|\lesssim \sqrt{\frac{s_1\log p}{n}}.%\lesssim\|D\Sigma_1-\hat D\hat\Sigma_1\|_F\lesssim \sqrt{\frac{s_1\log p}{n}}.
$$

For the term involving $\bm z$, when $\bm z\sim N_p(\bmu_1, \Sigma_1)$, we have \begin{align*}
&(\bm z-\bmu_1)^\top D(\bm z-\bmu_1)-(\bm z-\bmu_1)^\top \hat D(\bm z-\bmu_1)-(\tr(D\Sigma_1-\hat D \Sigma_1))\\
=&(\bm z-\bmu_1)^\top (\hat D-D)(\bm z-\bmu_1)-(\tr(D\Sigma_1-\hat D \Sigma_1))\\
\stackrel{d}{=}& \bm z_0^\top \Sigma_1^{1/2} (\hat D-D) \Sigma_1^{1/2} \bm z_0-\tr( \Sigma_1^{1/2} (\hat D-D) \Sigma_1^{1/2})\stackrel{def}{=}\sum_{i=1}^{p}\lambda_i (z_{0i}^2-1),
\end{align*}
where $\lambda_i$'s are the eigenvalues of $\Sigma_1^{1/2} (\hat D-D) \Sigma_1^{1/2}$. 
%(Remark: also, $\tr((\Sigma_1-\hat\Sigma_1)D)\le |\Sigma_1-\hat \Sigma_1|_\infty\sqrt s_1\|D\|_F$)

Since {with probability at least $1-p^{-1}$}, $$\sqrt{\sum_{i=1}^p\lambda_i^2}=\|\Sigma_1^{1/2} (\hat D-D) \Sigma_1^{1/2}\|_F\le \|\Sigma_1\|_2\|\hat D-D\|_F\lesssim\sqrt\frac{s_1\log p}{n},
$$ 
and {with probability at least $1-p^{-1}$},
$$\max_{i}|\lambda_i|\le\|\Sigma_1^{1/2} (\hat D-D) \Sigma_1^{1/2}\|_2\le \|\Sigma_1\|_2\|\hat D-D\|_2\lesssim\sqrt\frac{s_1\log p}{n},
$$
by Bernstein type inequality for sub-exponential random variables, see Vershynin (2011), we have for some $\tilde c_1>0$, $$
\Pro(|\sum_{i=1}^{p}\lambda_i (z_{0i}^2-1)|\ge t)\le 2\exp\{-\tilde c_1\min\{\frac{t^2}{s_1\log p/n},\frac{t}{\sqrt{s_1\log p/n}} \}\},
$$
which implies that for some $c_1>0$, \begin{align*}
\Pro(|\sum_{i=1}^{p}\lambda_i (z_{0i}^2-1)|\ge M\sqrt\frac{s_1\log p}{n})\le e^{-c_1M}+Cp^{-1}.
\end{align*} 

For $(\hat\bbeta-\bbeta)^\top\bm z$, when  $\bm z\sim N_p(\bmu_1, \Sigma_1)$, we have \begin{align*}
(\hat\bbeta-\bbeta)^\top\bm z\sim N((\hat\bbeta-\bbeta)^\top\bmu_1,(\hat\bbeta-\bbeta)^\top\Sigma_1(\hat\bbeta-\bbeta)).
\end{align*}
Since   {with probability at least $1-p^{-1}$},\begin{align*}
|(\hat\bbeta-\bbeta)^\top\bmu_1|\le\|\hat\bbeta-\bbeta\|_2\cdot\|\bmu_1\|_2\lesssim \sqrt\frac{s_2\log p}{n},
\end{align*}
and  {with probability at least $1-p^{-1}$},\begin{align*}
|(\hat\bbeta-\bbeta)^\top\Sigma_1(\hat\bbeta-\bbeta)|\le\|\Sigma_1\|_2\cdot\|\hat\bbeta-\bbeta\|_2^2\le \frac{s_2\log p}{n},
\end{align*}
we have for some $c_2>0$, $$
\Pro(|(\hat\bbeta-\bbeta)^\top\bm z|>M\sqrt\frac{s_2\log p}{n})\le e^{-c_2M^2}+{ Cp^{-1}}.
$$

Lastly, $$
|2\log(\frac{\pi_1}{\pi_2})-\log(\frac{\hat\pi_1}{\hat\pi_2})|\lesssim |\hat\pi_1-\pi_1|+|\hat\pi_2-\pi_2|.
$$
and by Hoeffding inequality, for $k\in[2]$, there are some constant $c_H>0$, such that 
$$
\Pro(|\hat\pi_k-\pi_k|>t)\le\exp(-c_H\cdot nt^2).
$$
We have for some constant $c,M_H>0$,
$$
\Pro(|2\log(\frac{\pi_1}{\pi_2})-\log(\frac{\hat\pi_1}{\hat\pi_2})|>M_H\sqrt{\frac{1}{n}})\le e^{-cM_H}.
$$
%{\cyan $$
%\Pro(|(\hat\bbeta-\bbeta)^\top\bm z|>t)\lesssim \exp(-\frac{ct^2}{s\log p/n})
%$$}

%When $\bm z\sim N(\bmu_2, \Sigma_2)$, let $\bm z_0=\bm z-\bmu_2$, then \begin{align*}
%(\bm z-\bmu_1)^\top D(\bm z-\bmu_1)=&(\bm z_0+\bmu_2-\bmu_1)^\top D(\bm z_0+\bmu_2-\bmu_1)\\
%=&\bm z_0^\top D\bm z_0+(\bmu_2-\bmu_1)^\top D\bm z_0+\bm z_0^\top D(\bmu_2-\bmu_1)+(\bmu_2-\bmu_1)^\top D(\bmu_2-\bmu_1)
%\end{align*}

Therefore,  there exists some $c>0$, such that for any $M>0$, $$
\Pro_{\bm z\sim N_p(\bmu_1,\Sigma_1)}(M(\bm z)>M \sqrt\frac{(s_1+s_2)\log p}{n})\le e^{-cM}+Cp^{-1}.
$$
%if we let $G(\bm z)=(\bm z-\bmu_1)^\top D(\bm z-\bmu_1)-2\bbeta^\top(\bm z-\bar \bmu)-\log(|D\Sigma_1+I_p|)$, $\hat G(\bm z)=(\bm z-\hat\bmu_1)^\top \hat D(\bm z-\hat \bmu_1)-2\hat \bbeta^\top(\bm z-\frac{\hat \bmu_1+\hat \bmu_2}{2})-\log(|\hat D\hat \Sigma_1+I_p|)$, and  $M(\bm z)=G(\bm z)-\hat G(\bm z)$,% we have for some $M>0$, $$
%\Pro(M(\bm z)>M\log n\sqrt\frac{(s_1+s_2)\log p}{n})\le\frac{1}{n}.
%$$

%Similar as before, and use the following fact
%\begin{align*}
%&|\tr(D\Sigma_1-\hat D \Sigma_1)-\tr(D\Sigma_2-\hat D \Sigma_2)|= |\tr((D-\hat D)(\Sigma_1-\Sigma_2))|\\
%\le&\|\hat D-D\|_F\cdot \|\Sigma_1-\Sigma_2\|_F\\
%\le&\sqrt{\frac{s_1\log p}{n}}  \|\Sigma_1D\Sigma_2\|_F\\
%\le&\sqrt{\frac{s_1\log p}{n}}  \|\Sigma_1\|_2\|D\|_F\|\Sigma_2\|_2\\
%\lesssim& \sqrt{\frac{s_1\log p}{n}}.
%\end{align*}

Then it follows that 
\begin{align*}
&R(\hat G_{\rm SDAR})-R_{\bth}(G^*_{\bth})\\
=&\frac{1}{2}\int_{Q(\bm z)>0} \frac{\pi_1}{(2\pi)^{p/2}|\Sigma_1|^{1/2}}e^{-1/2\cdot (\bm z-\bmu_1)^\top\Omega_1 (\bm z-\bmu_1)} d\bm z\\
&+\frac{1}{2}\int_{Q(\bm z)\le0} \frac{\pi_2}{(2\pi)^{p/2}|\Sigma_2|^{1/2}}e^{-1/2\cdot (\bm z-\bmu_2)^\top\Omega_2 (\bm z-\bmu_2)} d\bm z\\
&-\frac{1}{2}\int_{\hat Q(\bm z)>0} \frac{\pi_1}{(2\pi)^{p/2}|\Sigma_1|^{1/2}}e^{-1/2\cdot (\bm z-\bmu_1)^\top\Omega_1 (\bm z-\bmu_1)} d\bm z\\
&-\frac{1}{2}\int_{\hat Q(\bm z)\le0} \frac{\pi_2}{(2\pi)^{p/2}|\Sigma_2|^{1/2}}e^{-1/2\cdot (\bm z-\bmu_2)^\top\Omega_2 (\bm z-\bmu_2)} d\bm z.
\end{align*}
\begin{align*}
&R(\hat G_{\rm SDAR})-R_{\bth}(G^*_{\bth})\\
=&\frac{1}{2}\int_{Q(\bm z)>0} \frac{1}{(2\pi)^{p/2}}e^{-1/2\cdot (\bm z-\bmu_1)^\top\Omega_1 (\bm z-\bmu_1)-\log |\Sigma_1|/2+\log\pi_1} \\
&- \frac{1}{(2\pi)^{p/2}}e^{-1/2\cdot (\bm z-\bmu_2)^\top\Omega_2 (\bm z-\bmu_2)-\log |\Sigma_2|/2+\log\pi_2} d\bm z\\
&-\frac{1}{2}\int_{\hat Q(\bm z)>0} \frac{1}{(2\pi)^{p/2}}e^{-1/2\cdot (\bm z-\bmu_1)^\top\Omega_1 (\bm z-\bmu_1)-\log |\Sigma_1|/2+\log \pi_1}\\
&- \frac{1}{(2\pi)^{p/2}}e^{-1/2\cdot (\bm z-\bmu_2)^\top\Omega_2 (\bm z-\bmu_2)-\log |\Sigma_2|/2+\log \pi_2} d\bm z\\
=&\frac{1}{2}\int_{Q(\bm z)>0} \frac{1}{(2\pi)^{p/2}}e^{-1/2\cdot (\bm z-\bmu_1)^\top\Omega_1 (\bm z-\bmu_1)-\log |\Sigma_1|/2}(1-e^{-Q(\bm z)})d\bm z\\
&-\frac{1}{2}\int_{\hat Q(\bm z)>0} \frac{1}{(2\pi)^{p/2}}e^{-1/2\cdot (\bm z-\bmu_1)^\top\Omega_1 (\bm z-\bmu_1)-\log |\Sigma_1|/2} (1-e^{-Q(\bm z)}) d\bm z
\end{align*}
Then it follows
\begin{align*}
&R(\hat G_{\rm SDAR})-R_{\bth}(G^*_{\bth})\\
\le&\frac{1}{2}\int_{Q(\bm z)>0, \hat Q(\bm z)\le 0} \frac{1}{(2\pi)^{p/2}}e^{-1/2\cdot (\bm z-\bmu_1)^\top\Omega_1 (\bm z-\bmu_1)-\log |\Sigma_1|/2}(1-e^{-Q(\bm z)})d\bm z\\
=&\frac{1}{2}\int_{Q(\bm z)>0, Q(\bm z)\le Q(\bm z)-\hat Q(\bm z)} \frac{1}{(2\pi)^{p/2}}e^{-1/2\cdot (\bm z-\bmu_1)^\top\Omega_1 (\bm z-\bmu_1)-\log |\Sigma_1|/2}(1-e^{-Q(\bm z)})d\bm z\\
=&\frac{1}{2}\E_{\bm z\sim N_p(\bmu_1, \Sigma_1)}[(1-e^{-Q(\bm z)}) \1\{0<Q(\bm z)\le M(\bm z) \}]\\
%\le&\frac{1}{2}\E_{\bm z\sim N_p(\bmu_1, \Sigma_1)}[G(\bm z)\cdot \1\{0<G(\bm z)\le M(\bm z) \}]\\
%\le&\frac{1}{2}\E_{\bm z\sim N_p(\bmu_1, \Sigma_1)}[\max\{0,M(\bm z)\}]\\
%=&\frac{1}{2}\int_{0}^{+\infty} \Pro_{\bm z\sim N_p(\bmu_1, \Sigma_1)}(M(\bm z)>M)\: dM.
%\le&\frac{1}{2}\E_{\bm z\sim N_p(\bmu_1, \Sigma_1)}[\frac{M(\bm z)^2}{G(\bm z)}]\\
=&\frac{1}{2}\E_{\bm z\sim N_p(\bmu_1, \Sigma_1)}\left[(1-e^{-Q(\bm z)}) \1\{0<Q(\bm z)\le M(\bm z) \}\cdot \1\{M(\bm z)<M\log n \sqrt{\frac{(s_1+s_2)\log p}{n}} \}\right]\\
&+\frac{1}{2}\E_{\bm z\sim N_p(\bmu_1, \Sigma_1)}\left[(1-e^{-Q(\bm z)}) \1\{0<Q(\bm z)\le M(\bm z) \}\cdot \1\{M(\bm z)\ge M \log n \sqrt{\frac{(s_1+s_2)\log p}{n}} \}\right]\\
\le& \frac{1}{2}\E_{\bm z\sim N_p(\bmu_1, \Sigma_1)}\left[(1-e^{-Q(\bm z)}) \1\{0<Q(\bm z)\le M(\bm z) \}\cdot \1\{M(\bm z)<M\log n \sqrt{\frac{(s_1+s_2)\log p}{n}} \}\right]\\
&+\Pro_{\bm z\sim N_p(\bmu_1, \Sigma_1)}(M(\bm z)\ge M \log n \sqrt{\frac{(s_1+s_2)\log p}{n}}) \\
\lesssim& \E_{\bm z\sim N_p(\bmu_1, \Sigma_1)}\left[(1-e^{-Q(\bm z)}) \1\{0<Q(\bm z)\le M(\bm z) \}\cdot \1\{M(\bm z)<M\log n \sqrt{\frac{(s_1+s_2)\log p}{n}} \}\right]\\
&+n^{-1}+p^{-1} \\
\lesssim& \log n\cdot\sqrt\frac{(s_1+s_2)\log p}{n}\cdot\E_{\bm z\sim N_p(\bmu_1, \Sigma_1)}\left[ \1\{0<Q(\bm z)\le M\log n \sqrt{\frac{(s_1+s_2)\log p}{n}} \}\right]+n^{-1}+p^{-1} \\
\lesssim& \log^2n\cdot\frac{(s_1+s_2)\log p}{n},
\end{align*}   
where the last inequality uses the assumption that $\sup_{|x|<\delta}f_{Q,\bth}(x)< M_2$.

\begin{supplement}
%\sname{Supplement to}\label{suppA}
\stitle{Supplement to ``A Convex Optimization Approach to High-dimensional Sparse Quadratic Discriminant Analysis''.}
\slink[url]{http://www-stat.wharton.upenn.edu/$\sim$tcai/paper/SQDA-Supplement.pdf}
\sdescription{
The supplement provides a detailed proof of Theorem~\ref{sparse-lb}, which is the lower bound of the misclassification error for high-dimensional QDA problem with sparsity assumptions, and proofs of Theorem~\ref{para-copula} and \ref{Rn-copula}, the convergence rate of CSQDA under the Gaussian Copula Model. In addition, proofs of the technical lemmas used in the proofs of the main results are given.
}
\end{supplement}

\bibliographystyle{plainnat}
\bibliography{MyReference}

\end{document}